\documentclass[11pt,a4paper,twoside]{article}
\pdfoutput=1
\usepackage[T1]{fontenc}
\usepackage[utf8]{inputenc}
\usepackage{multicol}
\usepackage{amssymb}

\usepackage{nomencl}
\makenomenclature
\usepackage{times}

\usepackage{lipsum}
\usepackage{authblk}
\usepackage[top=2.5cm, bottom=2.5cm, left=2.5cm, right=2.5cm]{geometry}
\usepackage{fancyhdr}
\usepackage{adjustbox}
 \usepackage[round]{natbib}
\usepackage[colorlinks,bookmarksopen,bookmarksnumbered,citecolor=red,urlcolor=red]{hyperref}
\usepackage{algorithm}
\usepackage{algorithmic}
\usepackage{boxedminipage}
\usepackage{comment}
\usepackage{eqparbox}
\usepackage{multirow}
\usepackage{caption}
\usepackage{subcaption}

\usepackage{amsmath}
\usepackage{amsfonts}
\usepackage{graphicx}

\usepackage{tikz}

\usepackage{setspace}
\newlength{\commentindent}
\usepackage{array}
\newcolumntype{R}[1]{>{\raggedleft\let\newline\\\arraybackslash\hspace{0pt}}m{#1}}

\newcommand\BibTeX{{\rmfamily B\kern-.05em \textsc{i\kern-.025em b}\kern-.08em
T\kern-.1667em\lower.7ex\hbox{E}\kern-.125emX}}

\newcommand{\dive}{\nabla \cdot}
\newcommand{\grad}{\nabla}
\newcommand{\lap}{\Delta}

\usepackage{titlesec}
\titleformat*{\section}{\large\bfseries}
\titleformat*{\subsection}{\large\bfseries}

\let\OLDthebibliography\thebibliography
\renewcommand\thebibliography[1]{
	\OLDthebibliography{#1}
	\setlength{\parskip}{0pt}
	\setlength{\itemsep}{0pt plus 0.3ex}
}

\renewenvironment{abstract}{%
	\begin{center}
		\begin{adjustbox}{minipage=1.0\textwidth,center} 
			\rule{\textwidth}{0.0pt}}
		{\par\noindent\rule{\textwidth}{0.0pt}  
		\end{adjustbox}
	\end{center} 
}

\makeatletter

\renewcommand\@maketitle{%
	\begin{adjustbox}{minipage=1.0\textwidth}
		\begin{center}
			\vskip 1.0em
			\let\footnote\thanks 
			{\large \@title \par }
			\vskip 1.0em
			{\normalsize \@author \par}
		\end{center}
	\end{adjustbox}
	\vskip 2.0em \par
}
\makeatother

\usepackage{etoolbox}
\renewcommand\nomgroup[1]{%
  \item[\bfseries
  \ifstrequal{#1}{G}{Greek Symbols}{%
  \ifstrequal{#1}{S}{Symbols}{%
  \ifstrequal{#1}{A}{Abbreviations}{%
  \ifstrequal{#1}{B}{Subscripts}{%
  \ifstrequal{#1}{P}{Superscripts}{
  }}}}}]}

\usepackage{etoolbox}
\patchcmd{\subequations}{}%
{}{}{}

\begin{document}


\title{\textbf{Physics Informed Neural Networks for Two Dimensional Incompressible Thermal Convection Problems}}
\author{Atakan ~Aygun \footnote{Corresponding author.} }
\author{Ali ~Karakus}

\affil{Department of Mechanical Engineering, Middle East Technical University, Ankara, Turkey 06800 \\ atakana@metu.edu.tr, ORCID: 0000-0003-4399-1935 \\ akarakus@metu.edu.tr, ORCID: 0000-0002-9659-6712}
\date{\vspace{-5ex}}
\maketitle

\begin{abstract}

\textbf{Abstract} Physics informed neural networks (PINNs) have drawn attention in recent years in engineering problems due to their effectiveness and ability to tackle the problems without generating complex meshes. PINNs use automatic differentiation to evaluate differential operators in conservation laws and hence do not need to have a discretization scheme. Using this ability, PINNs satisfy governing laws of physics in the loss function without any training data. In this work, we solve various incompressible thermal convection problems including real applications and problems with comparable numerical or analytical results.  We first consider a channel problem with an analytical solution to show the accuracy of our network. Then, we solve a thermal convection problem in a closed enclosure in which the flow is only due to the temperature gradients on the boundaries. Lastly, we consider steady and unsteady partially blocked channel problems that resemble industrial applications to power electronics.


\textbf{\begin{center} 
{\large Sıkıştırılamaz Isıl Taşınım Problemlerinin Fizikle Öğrenen Yapay Sinir Ağları ile Çözümü }
\end{center}
}

\textbf{Ozet} Fizikle öğrenen sinir ağları (PINN'ler), etkinlikleri ve karmaşık ağlar oluşturmadan problemlerin üstesinden gelme yetenekleri nedeniyle son yıllarda mühendislik problemlerinde dikkat çekmiştir. PINN'ler, koruma yasalarında diferansiyel operatörleri uygulamak için otomatik türevlenmeyi kullanır ve bu nedenle bir ayrıklaştırma şemasına sahip olmaları gerekmeden ve herhangi bir eğitim verisi olmadan kayıp fonksiyonunda geçerli fizik yasalarını karşılamaya çalışır. Bu çalışmada, sayısal karşılaştırmalar ve gerçek hayata benzer uygulamalar dahil olmak üzere farklı sıkıştırılamaz ısıl taşınım problemleri üzerinde uygulamalar sunuyoruz. Önce analitik çözümü olan bir kanal problemini ağımızın doğruluğunu göstermek için ele alıyoruz. Ardından, akışın yalnızca sınırlardaki sıcaklık gradyanlarından kaynaklandığı kapalı bir mahfaza içindeki bir termal konveksiyon problemi çözülmüştür. Son olarak, endüstriyel uygulamaları güç elektroniğine benzeyen sabit ve kararsız kısmen bloke kanal problemleri ele alınmıştır.

\end{abstract}

\nomenclature[A]{DNS}{Direct Numerical Simulation}
\nomenclature[A]{NS}{Navier-Stokes}
\nomenclature[A]{PDE}{Partial Differential Equation}
\nomenclature[A]{PINN}{Physics Informed Neural Network}
\nomenclature[A]{\(Gr\)}{Grashof Number $[= g \beta \left(T-T_r\right)d^3/\nu^2]$}
\nomenclature[A]{\(Pr\)}{Prandtl Number $[= \nu / \alpha]$}
\nomenclature[A]{\(Re\)}{Reynolds Number $[= u_r d/ \nu]$}
\nomenclature[A]{\(Ra\)}{Rayleigh Number $[= GrPr]$}
\nomenclature[A]{MLP}{Multilayer Perceptron}
\nomenclature[A]{NTK}{Neural Tangent Kernel}
\nomenclature[A]{FCN}{Fully Connected Network}
\nomenclature[A]{DGM}{Deep Galerkin Method}

\nomenclature[B]{\(r\)}{Reference Value}
\nomenclature[B]{\(\theta\)}{Temperature Related Value }
\nomenclature[B]{\(\mathbf{u}\)}{Velocity Related Value }
\nomenclature[B]{\(D\)}{Dirichlet Boundary}
\nomenclature[B]{\(N\)}{Neumann Boundary}
\nomenclature[B]{\(R\)}{Residual}
\nomenclature[B]{\(BC\)}{Boundary Condition}
\nomenclature[B]{\(IC\)}{Initial Condition}

\nomenclature[S]{\(t\)}{Non-dimensional Time}
\nomenclature[S]{\(\mathbf{u}\)}{Velocity Vector}
\nomenclature[S]{\(\mathbf{s}_\mathbf{u}\)}{Source Term of Momentum Equation}
\nomenclature[S]{\(s_\theta\)}{Source Term of Energy Equation}
\nomenclature[S]{\(g_D\)}{Dirichlet Boundary Condition}
\nomenclature[S]{\(g_N\)}{Neumann Boundary Condition}
\nomenclature[S]{\(p\)}{Non-dimensional Pressure}
\nomenclature[S]{\(\mathcal{N}\)}{Generalized Differential Operator}
\nomenclature[S]{\(\mathcal{L}\)}{Loss Term of a Neural Network}
\nomenclature[S]{\(L_2\)}{$L_2$ Vector Norm of Error}

\nomenclature[G]{\(\nu\)}{Kinematic Viscosity $ [= m^2/s]$}
\nomenclature[G]{\(\alpha\)}{Thermal Diffusivity $ [= m^2/s]$}
\nomenclature[G]{\(\theta\)}{Non-dimensional Temperature}
\nomenclature[G]{\(\beta\)}{Expansion Coefficient $ [= 1/^{\circ}C]$}
\nomenclature[G]{\(\grad\)}{Gradient Operator}
\nomenclature[G]{\(\lap\)}{Laplace Operator}
\nomenclature[G]{\(\dive\)}{Divergence Operator}
\nomenclature[G]{\(\sigma\)}{Activation Function}
\nomenclature[G]{\(\omega\)}{Weight of Loss Terms}
\printnomenclature

\section*{INTRODUCTION}
Thermal convection problems arise in many practical engineering applications such as cooling of electronic chips. This type of real life analysis of fluid flow and heat transfer require high degrees of freedom to minimize the numerical error. This can be achieved with high quality mesh or high order discretizations. However, mesh generation is time consuming and need expertise and high order simulation tools for this type of problems are computationally demanding.

The incompressible thermal convection is studied in the literature with various numerical methods \citep{bairi2014review}. \cite{tang1993least} used a least squares finite element method based on a velocity, pressure, vorticity, temperature, heat flux formulation for time dependent problems. \cite{hossain2021spectral} developed a spectral/hp element method for the Direct Numerical Simulation (DNS) of incompressible thermal convective flows by considering Boussinesq type thermal body-forcing with periodic boundary conditions and enforcing a constant volumetric flow rate. In \cite{karakus_accelerated_2022}, the author presented a GPU accelerated nodal discontinuous Galerkin method on unstructured triangular meshes for solving problems on different convective regimes.

Apart from the traditional numerical methods such as finite difference, finite volume and finite element methods, data driven machine learning methods are used to solve the partial differential equations (PDE) \citep{willard2020physicsMachineLearningSurvey}. These regression methods offer effective and mesh free approaches \citep{karniadakis_physics-informed_2021}. Neural networks were employed first to solve the PDEs as in \cite{lee_neural_1990} and \cite{lagaris_artificial_1998}, and in \cite{raissi_inferring_2017} and \cite{raissi_machine_2017} the authors employed Gaussian processes regression to accurately predict the solution and provide the uncertainty in the model. \cite{raissi_physics-informed_2019} introduced the concept of physics informed neural networks (PINNs) that use automatic differentiation \citep{baydin_automatic_2017} to solve forward and inverse problems for several types of PDEs. 

To solve PDEs using PINNs, generally fully connected networks are used. However, for different type of problems, plain fully connected networks perform poorly. In the learning process, these networks have a learning bias towards low frequency functions that is called spectral bias \citep{rahaman2019spectral}. This reduces the accuracy in which the target function exhibit high frequency or multi-scale behavior. To overcome this problem, \cite{tancik2020fourier} and \cite{wang2021eigenvector} proposed Fourier feature mapping of the input vectors before feeding them into the network. The input coordinates \( \mathbf{v}\) is mapped with a function \( \gamma(\mathbf{v}) = [a_1 \cos(2\pi\mathbf{b}_1^T\mathbf{v}), a_1 \sin(2\pi\mathbf{b}_1^T\mathbf{v}), \dots, a_m \cos(2\pi\mathbf{b}_m^T\mathbf{v}), a_m \sin(2\pi\mathbf{b}_m^T\mathbf{v})]\) and then passed into the multi layer perceptron (MLP). This mapping transforms the Neural Tangent Kernel (NTK) \citep{jacot2018neuraltangentkernel} into a stationary kernel, and enables controlling the learning of the range of frequencies by modifying the frequency vectors \( \mathbf{b}\) in the mapping function. \cite{tancik2020fourier} shows the performance of this method in many low-dimensional tasks in computer vision and \cite{wang2021eigenvector} shows its efficiency for challenging problems involving partial differential equations with multi-scale behavior where conventional PINN models might have issues, such as wave propagation and reaction-diffusion equations.

In another approach, \cite{esmaeilzadeh2020meshfreeflownet}, proposed a PDE constrained deep learning algorithm that reconstructs high resolution solutions using the low resolution physical solutions in space and time and solved the well known Rayleigh-Bénard instability problem. This model is referred as super-resolution model and enables effectively scaling to large domains and having physically valid solutions by regularizing the outputs with PDE constraints. The framework is named as MeshfreeFlowNet and consists of two subnetworks. The first network, named as Context Generation Network, is a convolutional encoder that takes low resolution physical input and creates a Latent Context Grid. This grid contains latent context vectors along with the spatio-temporal coordinates and these values are fed into another network called Continuous Decoding network modeled as a MLP. This framework allows the output to be continuous instead of a discrete output therefore removing the output resolution limitations. In addition, this continuous output allows to effectively compute the gradients of the output, therefore enables to enforce the PDE based constraints.

Due to the popular deep learning frameworks such as TensorFlow \citep{abadi_tensorflow_2016} and PyTorch \citep{paszke_pytorch_2019}, and their easy implementation, PINNs become quite popular for solving PDEs. It is used for solving incompressible and compressible Navier-Stokes equations \citep{rao_physics-informed_2020,jin_nsfnets_2021, cai_physics-informed_fluid_review_2022}, as well as in heat transfer problems \citep{cai_physics-informed_heat_2021}. There are also software libraries specifically designed for physics informed machine learning such as DeepXDE \citep{lu_deepxde_2021} and NeuralPDE \citep{zubov_neuralpde_2021}.

PINNs do not require mesh generation. Instead, the PDEs and any other constraints can be directly enforced into the loss function of the neural network using automatic differentiation by forcing the prediction to the target value. This includes zero residuals for conservation laws and satisfying the boundary/initial conditions.
In this work, we focus on the application of PINNs to coupled fluid flow and heat transfer problems. We present and validate our approach for different thermal convection regimes. The remainder of this paper is organized as follows. First, we present the mathematical formulation of incompressible Navier-Stokes equations coupled with the energy equation through Boussinesq approximation. Then we give brief information about the physics informed neural networks and their loss functions which is followed by the 2D numerical validation cases. The last section is about concluding remarks and future works.

\section*{FORMULATION}
We consider a two-dimensional domain $\Omega \subset \mathbb{R}^2$ and denote the boundary of $\Omega$ by $\partial\Omega$. Following the notation presented in \cite{karakus_gpu_2019}, we denote the Dirichlet and Neumann boundary conditions as $\partial\Omega_D$ and $\partial\Omega_N$ respectively on $\partial\Omega$. We are interested in the approximation of non-isothermal incompressible Navier-Stokes equations coupled  by the energy equation through Boussinesq approximation which reads
\begin{subequations}
\label{eq:INS1}
\begin{align}
  \dive\mathbf{u} &= 0 \quad &\text{in}\;\Omega\times (0,T] \label{eq:INS1_1}\\
  \frac{\partial \mathbf{u}}{\partial t} + \left(\mathbf{u}\cdot\grad\right)\mathbf{u} &= -\grad p+ \frac{1}{Re}\lap \mathbf{u} +\mathbf{s_u} \quad &\text{in}\;\Omega\times (0,T]\label{eq:INS1_2}, \\
 \frac{\partial \theta} {\partial t} + \left(\mathbf{u}\cdot\grad\right) \theta &=  \frac{1}{Re Pr} \lap  \theta +\mathbf{s_\theta} \quad &\text{in}\;\Omega\times (0,T],\label{eq:INS1_3}
\end{align}
\end{subequations}
in non-dimensional form and subject to the initial conditions 
\begin{equation}
\label{eq:initialCondition}
\mathbf{u} = \mathbf{u}_0, \quad \mathbf{\theta} = \mathbf{\theta}_0 \quad\text{for}\;t =0, \mathbf{x} \in \Omega, 
\end{equation}
and the boundary conditions
\begin{subequations}
\label{eq:BoundaryConditions}
\begin{align}\label{eq:DirichletBC_u}
\mathbf{u} &= \mathbf{g}_D\;\; \text{on}\;\;\mathbf{x}\in\partial\Omega_D^\mathbf{u}, t\in (0,T],\\
\label{eq:NeumannBC_u} \frac{\partial\mathbf{u}}{\partial \mathbf{n}} &= 0,\;p = 0 \;\; \text{on}\;\;\mathbf{x}\in\partial\Omega_N^\mathbf{u}, t\in (0,T].\\
\label{eq:DirichletBC_t}
\theta &= g_D\;\;\text{on}\;\;\mathbf{x}\in\partial\Omega_D^\theta, t\in (0,T],\\
\label{eq:NeumannBC_t}
\frac{\partial\theta}{\partial \mathbf{n}} &= g_N  \;\; \text{on}\;\;\mathbf{x}\in\partial\Omega_N^\theta, t\in (0,T].
\end{align}
\end{subequations}
Here  $\mathbf{u}$, $p$ and  $\theta$ are non-dimensional velocity,  static pressure and temperature fields, respectively. To get the nondimensional representation of the equation, following approach is used.
\begin{equation}
\begin{aligned}
    &x = \frac{x^*}{L_r}, \: t = \frac{t^*}{L_r/U_r}, \: u = \frac{u^*}{U_r}, \: p = \frac{p^*}{\rho_r U_r^2}, \\ &\rho = \frac{\rho^*}{\rho_r}, \: \nu = \frac{\nu^*}{\nu_r}, \: \alpha = \frac{\alpha^*}{\alpha_r}, \: \theta = \frac{T - T_r}{T_s}
\end{aligned}
\end{equation}
\noindent
Here the superscript * denotes the dimensional parameters and subscript \(r\) denotes the reference values for specific terms with reference length scale $L_r$, velocity $U_r$, density $\rho_r$, viscosity $\nu_r$, thermal diffusivity $\alpha_r$, and temperature $T_r$. Also, $Re$ is the Reynolds number and $Pr$ is the Prandtl number. $\mathbf{s}_\mathbf{u} =\left( \mathbf{g}\beta \left(T -T_r \right)L_r/\mathbf{u}_r \right)\theta$ is the forcing term for Navier-Stokes, where $\mathbf{g}$ is the gravitational acceleration, $\beta$ is the expansion coefficient and subscript $r$ refers the reference value for the corresponding field. $s_\theta = s_\theta\left(\theta, \grad \theta,\mathbf{u}\right)$ is the generic generation term for the energy equation written in terms of temperature. The superscripts $\mathbf{u}$ and $\theta$ in boundary representation separate the Dirichlet and Neumann conditions, represented with the subscripts $D$ and $N$, on the physical boundary set for flow and heat transfer equations.

\section*{PHYSICS INFORMED NEURAL NETWORKS}
\label{sec:PINN}
Consider a general partial differential equation expressed as:
\begin{subequations}
\label{eq:PDEsystem}
\begin{align}
 &\mathbf{u}_t + \mathcal{N}[u] = 0, \quad &\mathbf{x} \in \Omega,\; t\in[0,T]\label{eq:PDE}\\
 &\mathbf{u}(\mathbf{x},0) = f(\mathbf{x}), \quad &\mathbf{x} \in \Omega\label{eq:PDE IC}\\
 &\mathbf{u}(\mathbf{x},t) = g(\mathbf{x},t), \quad &\mathbf{x} \in \partial\Omega, \; t \in [0,T]\label{eq:PDE BC}
\end{align}
\end{subequations}
where $\mathcal{N}$ is a generalized differential operator that can be linear or nonlinear, $\mathbf{x} \in \mathbb{R}^d$ and $t$ are the spatial and temporal coordinates. $\Omega \;\text{and}\; \partial\Omega$ represent the computational domain and the boundary respectively. $\mathbf{u}(\mathbf{x},t)$ is the general solution of the PDE with $f(\mathbf{x})$ is the initial condition and $g(\mathbf{x},t)$ is the boundary condition.

The solution $\mathbf{u}(\mathbf{x},t)$ can be approximated by a fully connected network according to the framework of physics informed neural networks (PINN) proposed by \cite{raissi_physics-informed_2019}. This network takes the spatiotemporal coordinates $(\mathbf{x},t)$ as input and outputs a solution $\mathbf{u}_{NN}(\mathbf{x},t)$. Between these input and output layers, there exists multiple hidden layers. Each hidden layer takes input $X=[x_1,x_2,\dots,x_i]$, and outputs $Y=[y_1,y_2,\dots,y_j]$ through a nonlinear activation function $\sigma(\cdot)$ such as:
\begin{equation}\label{eq:ActivationFunc}
    y_j = \sigma(w_{i,j}x_j + b_j),
\end{equation}
where $w_{i,j}\;\text{and}\;b_j$ are trainable hyperparameters, weights and biases, respectively. These hyperparameters are tuned such that it minimizes a composite loss function in the form
\begin{equation}\label{eq:CompositeLossFunction}
    \mathcal{L} = \omega_D\mathcal{L}_D + \omega_R\mathcal{L}_R + \omega_{BC}\mathcal{L}_{BC} + \omega_{IC}\mathcal{L}_{IC},
\end{equation}
where
\begin{subequations}
\label{eq:LossFunctions}
\begin{align}
 \mathcal{L}_D &= \frac{1}{N_D}\sum_{i=1}^{N_D}|\mathbf{u}(\mathbf{x}^i,t^i) - \mathbf{u}^i|^2\\
 \mathcal{L}_R &= \frac{1}{N_R}\sum_{i=1}^{N_R}|\mathbf{u}_t + \mathcal{N}[\mathbf{u}(\mathbf{x}^i,t^i)]|^2\\
 \mathcal{L}_{BC} &= \frac{1}{N_{BC}}\sum_{i=1}^{N_{BC}}|\mathbf{u}(\mathbf{x}^i,t^i) - g(\mathbf{x}^i,t^i)|^2\\
 \mathcal{L}_{IC} &= \frac{1}{N_{IC}}\sum_{i=1}^{I_{BC}}|\mathbf{u}(\mathbf{x}^i,0) - f(\mathbf{x}^i)|^2.
\end{align}
\end{subequations}
$\mathcal{L}_D$ is a loss term indicating the accuracy of the prediction if there are some data available in the domain, $\mathcal{L}_R, \;\mathcal{L}_{BC}, \;\text{and}\; \mathcal{L}_{IC}$ represent the residual of the governing PDE, the boundary conditions, and the initial condition respectively. $N_D,\;N_R,\;N_{BC},\; \text{and}\;N_{IC}$ are the number of data points for different terms, $\omega_D,\; \omega_R,\; \omega_{BC} \;\text{and}\; \omega_{IC}$ are user specified weighting coefficients of each loss term.
To calculate the residuals for $\mathcal{L}_R$, one need to take the derivative of the output $\mathbf{u}$ with respect to inputs $(\mathbf{x},t)$. In PINNs, this can be achived by using automatic differentiation \citep{baydin_automatic_2017}. Automatic differentiation applies chain rule repeatedly to the elementary functions and arithmetic operations to achieve the derivative of the overall composition. It plays a key role in the development of PINNs by enabling the computation of the residual of the governing differential equation \citep{raissi_physics-informed_2019}. Automatic differentiation is well implemented in most deep learning frameworks such as TensorFlow \citep{abadi_tensorflow_2016} and PyTorch \citep{paszke_pytorch_2019}.

\begin{figure}[h]
    \centering
    \includegraphics[width=\linewidth]{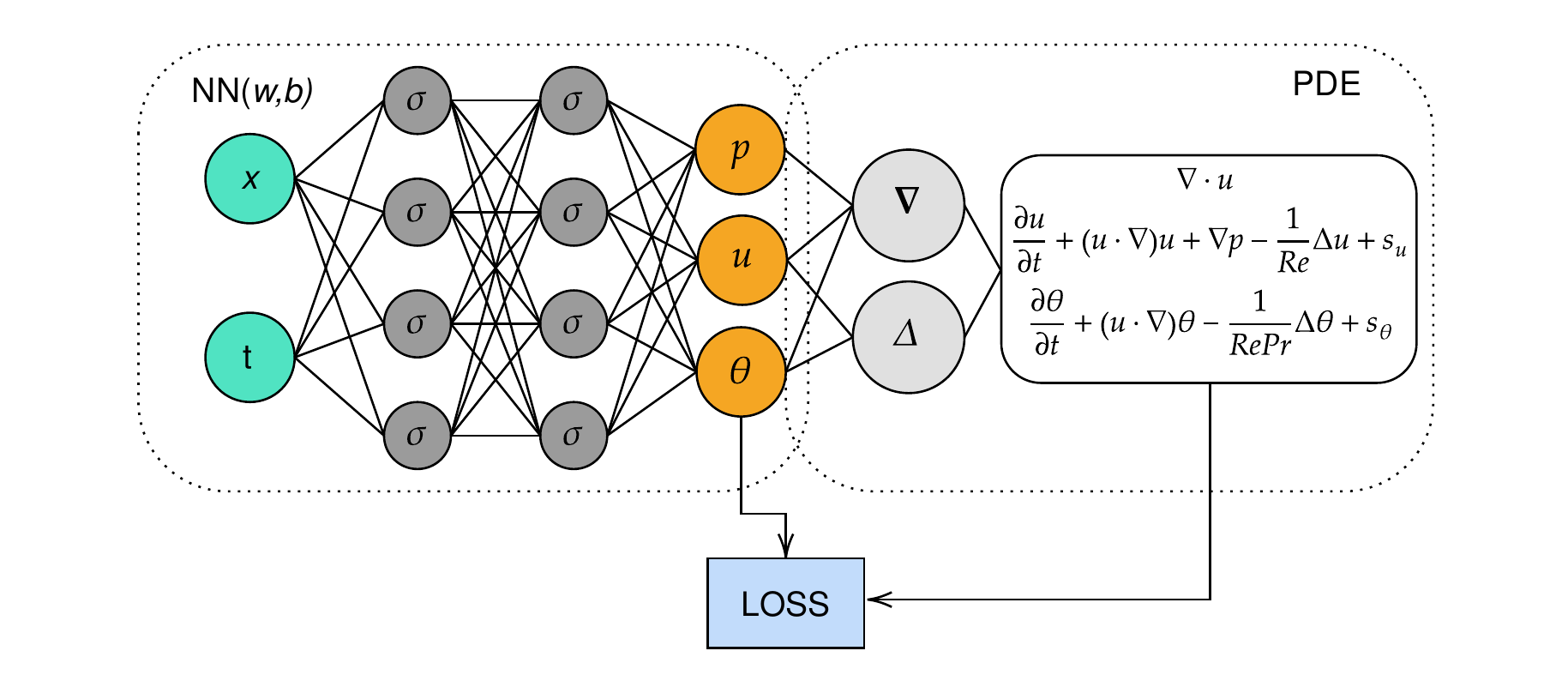}
    \caption{Schematic of a PINN framework}
    \label{fig:PINN}
\end{figure}

Figure \ref{fig:PINN} illustrates a schematic of the PINN framework in which the residuals of the coupled Navier-Stokes and energy equaitons are shown as the loss terms. This general schematic of the neural network takes the spatio-temporal coordinates as the input, and through the hidden layers it outputs the pressure, velocity, and temperature fields. After this output, the framework calculates the boundary losses from the boundary conditions and the data loss if there are any observations. In the next step, PINN uses automatic differentiation and calculates the residual inside the domain by enforcing the neural network output to the Navier-Stokes and energy equations. Also, if the boundary conditions are Neumann type, the neural network output can be differentiated to find the boundary loss. Then the overall loss can be calculated by adding the residual loss from the PDE, the boundary loss and if there exists any available observations, the data loss. This combined loss function is minimized by an optimization algorithm via changing the hyperparameters.

\section*{RESULTS}
\label{Results}
We have implemented our physics informed neural network on top of the NVIDIA Modulus framework \citep{hennigh2021nvidia}. We use the Adam optimizer \citep{kingma_adam_2017} to minimize the loss function defined in Equation \ref{eq:CompositeLossFunction}, and use 8 hidden layers with 40 units for the each test case where the parameters of the neural network initialized using the Glorot scheme \citep{glorot_understanding_2010}. We solve different 2D thermal convection tests to show the solutions by representing the velocity, pressure and the temperature fields. 

\subsection*{Poiseuille Flow}
\label{sec.Poiseuille Flow}
\begin{figure*}[ht!]
	\begin{center}
      \begin{subfigure}[b]{0.32\textwidth}
			\includegraphics[width=\textwidth]{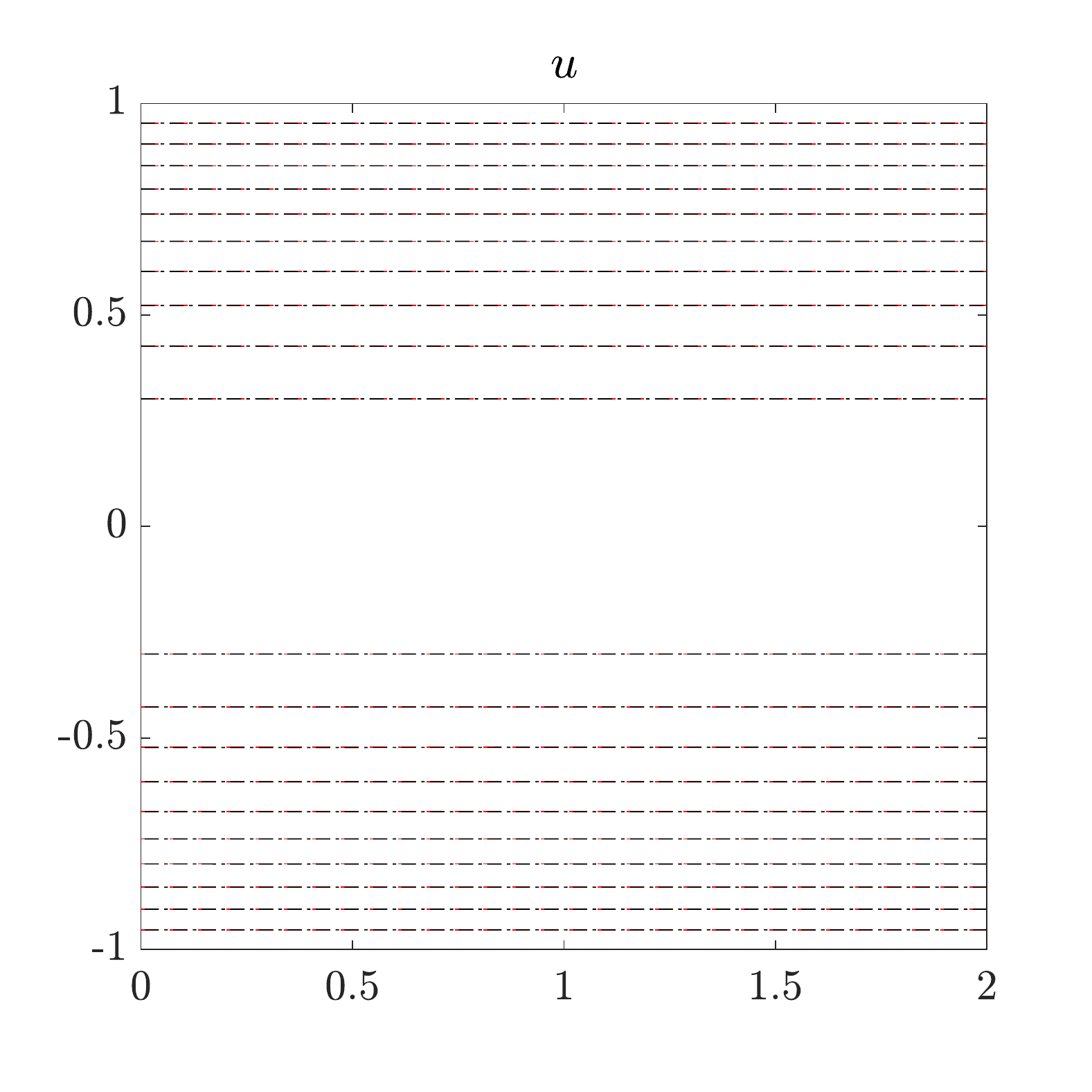} 
		\end{subfigure}
        ~
        \begin{subfigure}[b]{0.32\textwidth}
          \includegraphics[width=\textwidth]{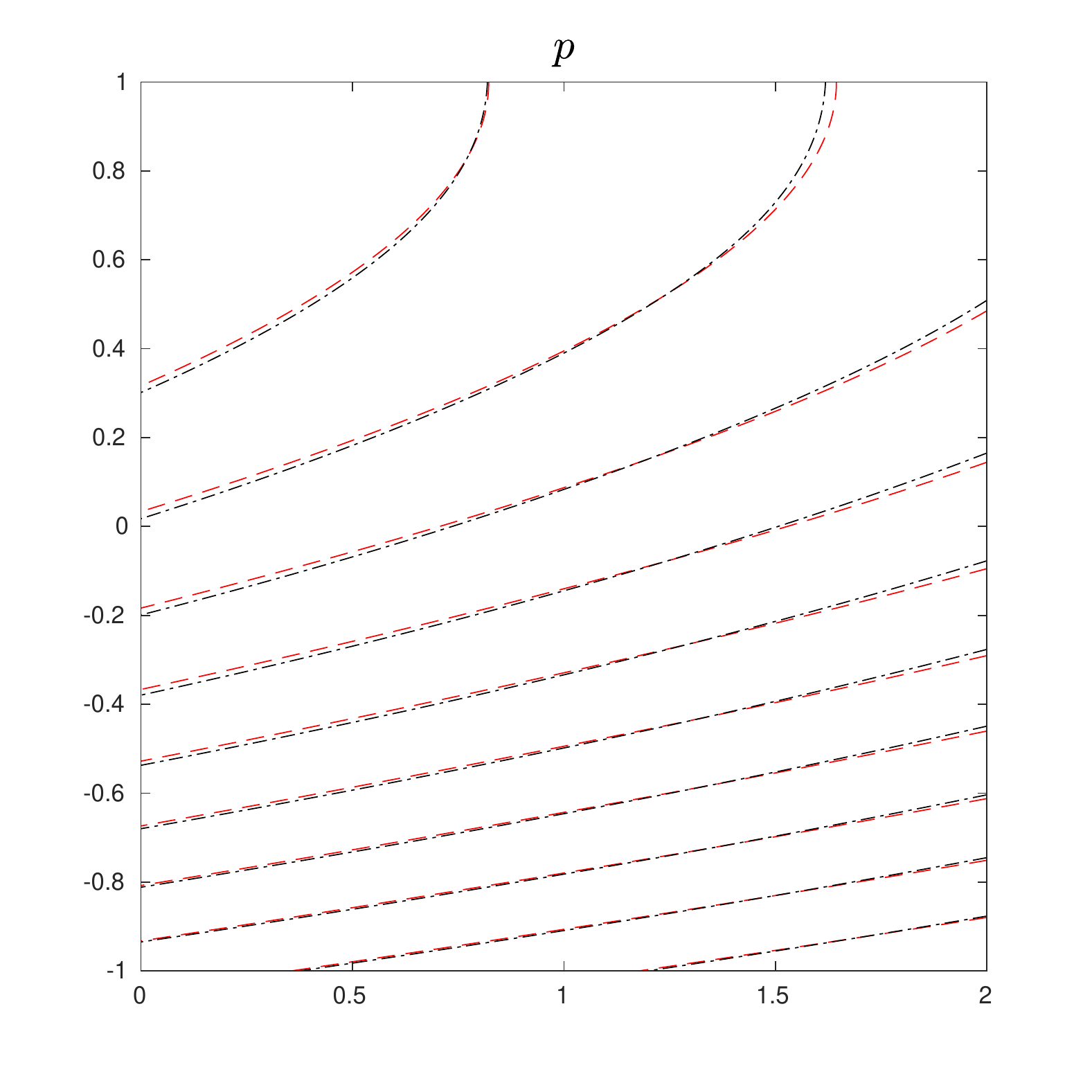}
      \end{subfigure}
        ~
        \begin{subfigure}[b]{0.32\textwidth}
          \includegraphics[width=\textwidth]{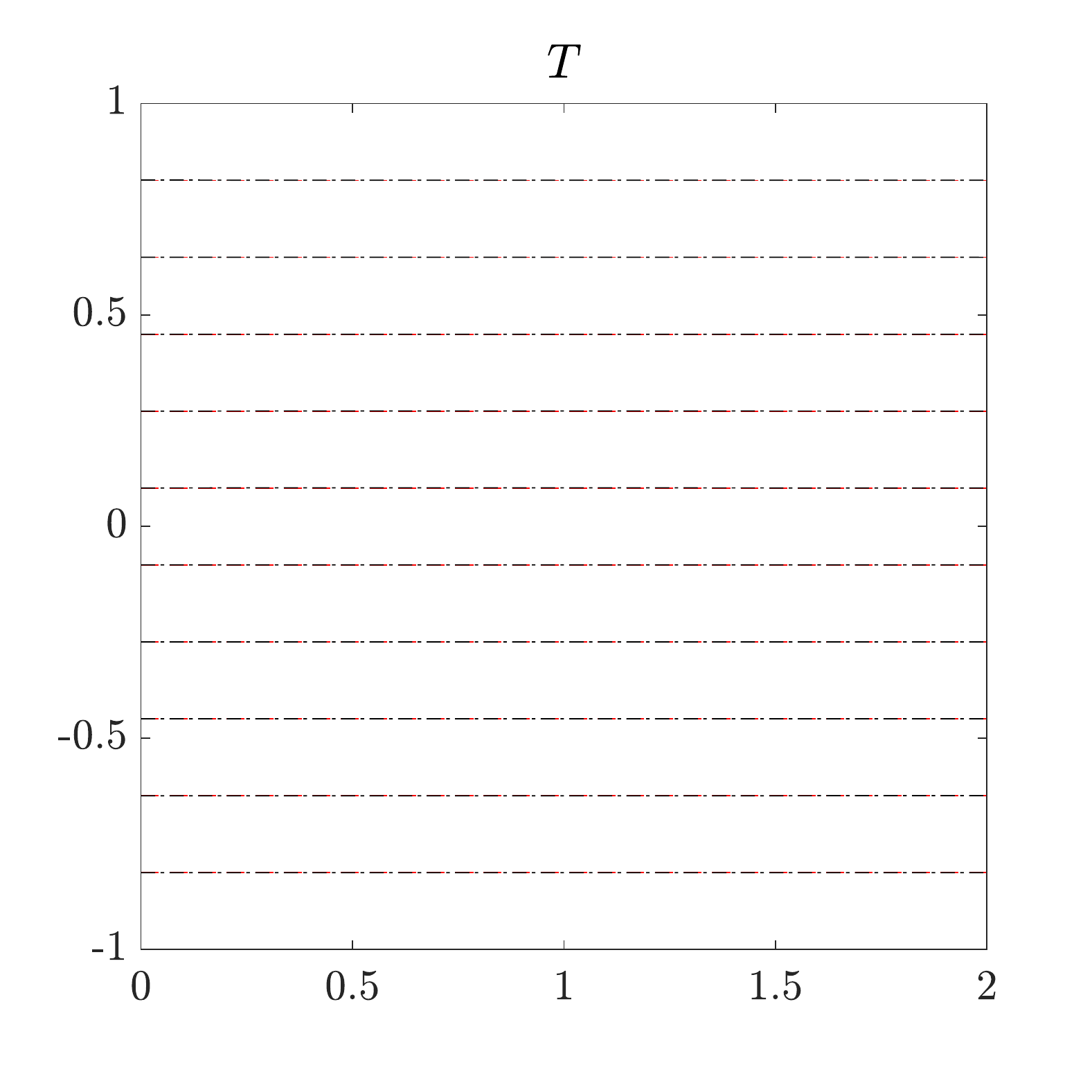}
      \end{subfigure}
	\end{center}
	\caption{Prediction of the Poiseuille flow with PINN. The $u$ velocity, pressure and the temperature fields are shown in order. Black contours show the exact solution while the red dashed contours show the solution with PINN.}
	\label{fig:channel_predict}
\end{figure*}

In the first test case, we consider two-dimensional channel flow with a fully developed Poiseuille profile. The channel dimension is $[0,2] \times [-1,1]$. The upper and lower walls have constant temperature of $\theta_L = 1$ and $\theta_U=0$. No-slip boundary conditions are imposed for upper and lower walls. The fully developed solution of the velocity field with linear temperature profile shown below is implemented as the boundary conditions of the inlet and the outlet. The flow conditions are stated as $Ra=10^3$, $Pr=0.71$ and $Re=100$.
\begin{align*}
    u = 1-y^2, \quad v = 0, \quad p = \frac{Ra}{2 Pr Re^2}\left(y- \frac{y^2}{2}\right)-\frac{2x}{Re}, \quad \theta=\frac{1-y}{2}
\end{align*}
 We trained our framework with 250 samples inside the domain and 30 samples on each boundary with 10000 iterations for this case. The training points are sampled using latin hypercube sampling, and the loss function for this problem contains only the Dirichlet boundary condition loss for the velocity and the temperature on the walls combined with the residual loss inside the domain. After training, we performed a prediction on a $(251\times251)$ grid and obtained the velocity, pressure, and temperature fields. The predicted fields can be seen in Figure \ref{fig:channel_predict}. The accuracy of the PINN is highly dependent on the weights of the loss function. In this case we tried different weights of the different terms of the loss function to match our solution with the exact solution. Especially for an accurate pressure field we increased the weights of the boundary condition losses. The solution in Figure \ref{fig:channel_predict} is obtained with a boundary loss weight $\omega_{BC}$ which is eight times higher than the weight of the residual loss $\omega_R$. Since the convective effects are not very dominant for this problem, boundary losses are dominant, so increasing the boundary loss weights increases the accuracy.
%

\begin{table*}[htb!]
	\caption{$L_2$ norm of the error of the predicted $u$ velocity and the temperature fields.}
	\centering
	\label{table:channel_observation_norm}
	\begin{tabular}{ c c c}
	\hline \hline
		Number of Observations & $u$ & $T$ \\ \hline \hline
		20  & 0.527 & 0.087 \\
		50  & 0.253 & 0.057 \\
		100 & 0.166 & 0.034 \\
		150 & 0.141 & 0.032 \\
    \hline \hline	
	\end{tabular}
\end{table*}

We test the performance of the PINNs with the addition of true observations at random points on the domain. We fused different number of randomly sampled exact solutions inside the domain to the network and add a data loss term into the loss function. The training process is done with 250 points inside the domain and 30 boundary points on each boundary besides the true solution points. In Table \ref{table:channel_observation_norm}, we can see the $L_2$ norm of the error of the predicted $u$ velocity and temperature fields. The number of observation represents the addition of the true solutions and increasing this number reduces the $L_2$ norm of the prediction of the velocity and the temperature from the true solution.
\subsection*{Differentially Heated Square Cavity}
\label{sec:Square_Cavity}
We focus on the natural convection problem on a two dimensional closed enclosure. The enclosure is a square cavity with its height denoted as $H=1$, and width as $W=1$. The boundary conditions of cavity are simple no-slip walls, $u=0,\;v=0$, on all four walls. The thermal boundary conditions on the left and right walls are prescribed as 
\begin{align*}
    \theta_L = 1, \quad \theta_R=0,
\end{align*}
and the upper and the lower walls are thermally insulated
\begin{align*}
    \frac{\partial\theta}{\partial y} = 0, \quad \text{for} \quad y=0 \quad \text{and} \quad y=H. 
\end{align*}
The flow conditions are $Pr=0.71$, and three different Rayleigh numbers as $Ra = 10^3,\;10^4,\;10^5$.

For the PINN solution, we sampled 150 points on each boundary, and 1000 collocation points inside the domain for the training process. Boundary points are used to minimize the loss of Dirichlet and Neumann boundary conditions, and collocation points are used to minimize the residual inside the domain. Automatic differentiation is used to calculate the derivatives on Neumann boundaries. 

\begin{table*}[htb!]
	\caption{Maximum and minimum velocities along the center lines of square cavity for $Pr=0.71$ and $Ra=10^3, 10^4, 10^5$. }
	\centering
	\label{table:square_cavity_1}
	\begin{tabular}{ c c c c c c c c}
		\hline \hline
		&       &  \multicolumn{2}{ c }{$ Ra=10^3$} & \multicolumn{2}{c}{$Ra=10^4$} &  \multicolumn{2}{c}{$Ra=10^5$} \\ \cline{3-4} \cline{5-6}  \cline{6-8} 
		&  &  $u_{max}$  &  $v_{max}$       &   $u_{max}$  &  $v_{max}$  &   $u_{max}$  &  $v_{max}$        \\  \hline\hline
	 PINN &     & $0.137 $ & $0.138$ &$0.192$ & $0.233$    & $ 0.128$ &  $0.258$ \\
	 \cite{karakus_accelerated_2022} &     & $0.137 $ & $0.139$ &$0.192$ & $0.233$    & $ 0.129$ &  $0.257$     \\
     \cite{stokos_development_2015}&     & $0.137 $ & $ 0.139$ &$0.192$ & $0.233$    & $0.130$ &  $0.256$     \\
	 \cite{de_vahl_davis_natural_1983} &     & $0.136 $ & $ 0.138$ &$0.192$ & $0.234$    & $ 0.153$ &  $0.261$     \\
     \hline \hline	
	\end{tabular}
\end{table*}

\begin{figure*}[ht!]
	\begin{center}
      \begin{subfigure}[b]{0.32\textwidth}
			\includegraphics[width=\textwidth]{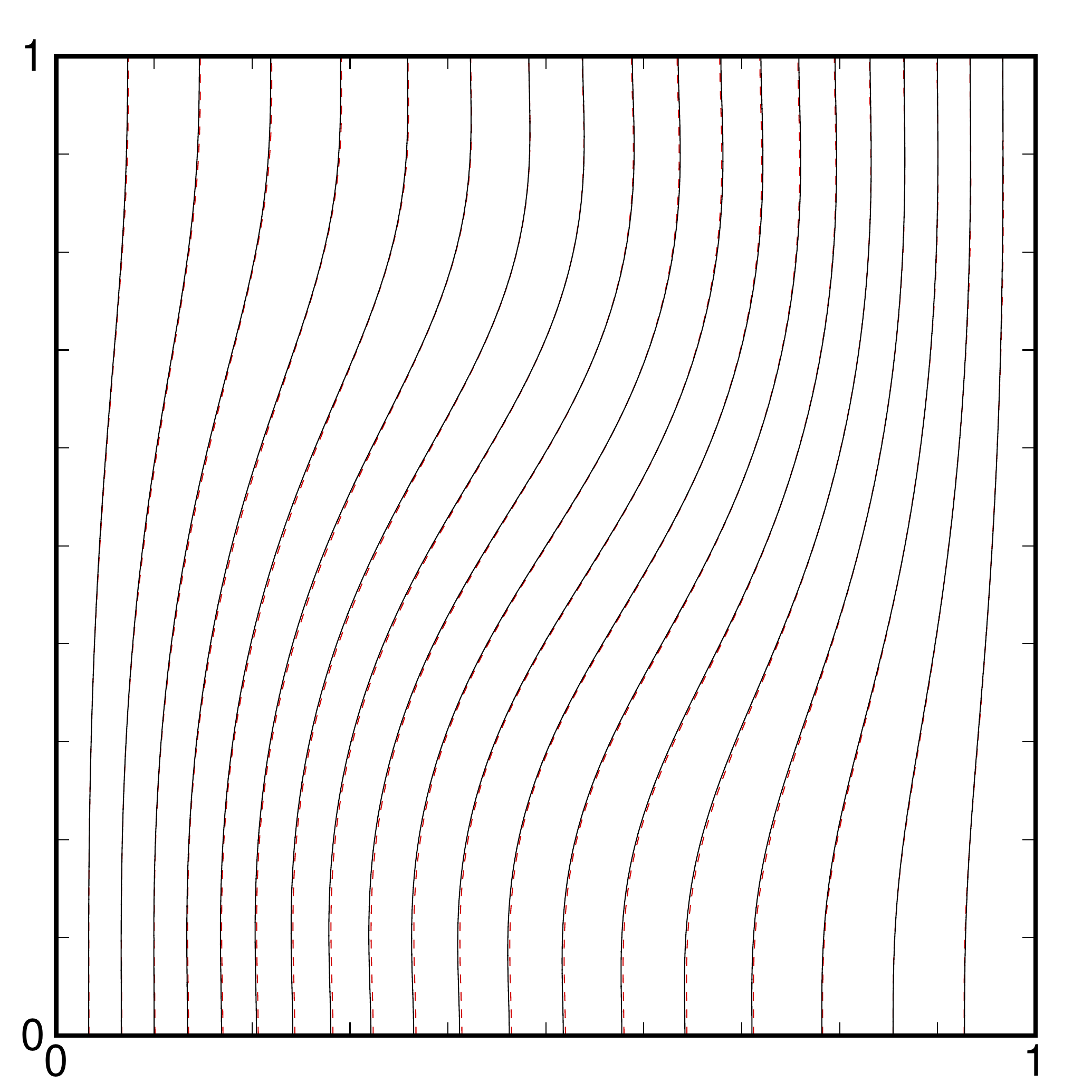} 
		\end{subfigure}
        ~
        \begin{subfigure}[b]{0.32\textwidth}
          \includegraphics[width=\textwidth]{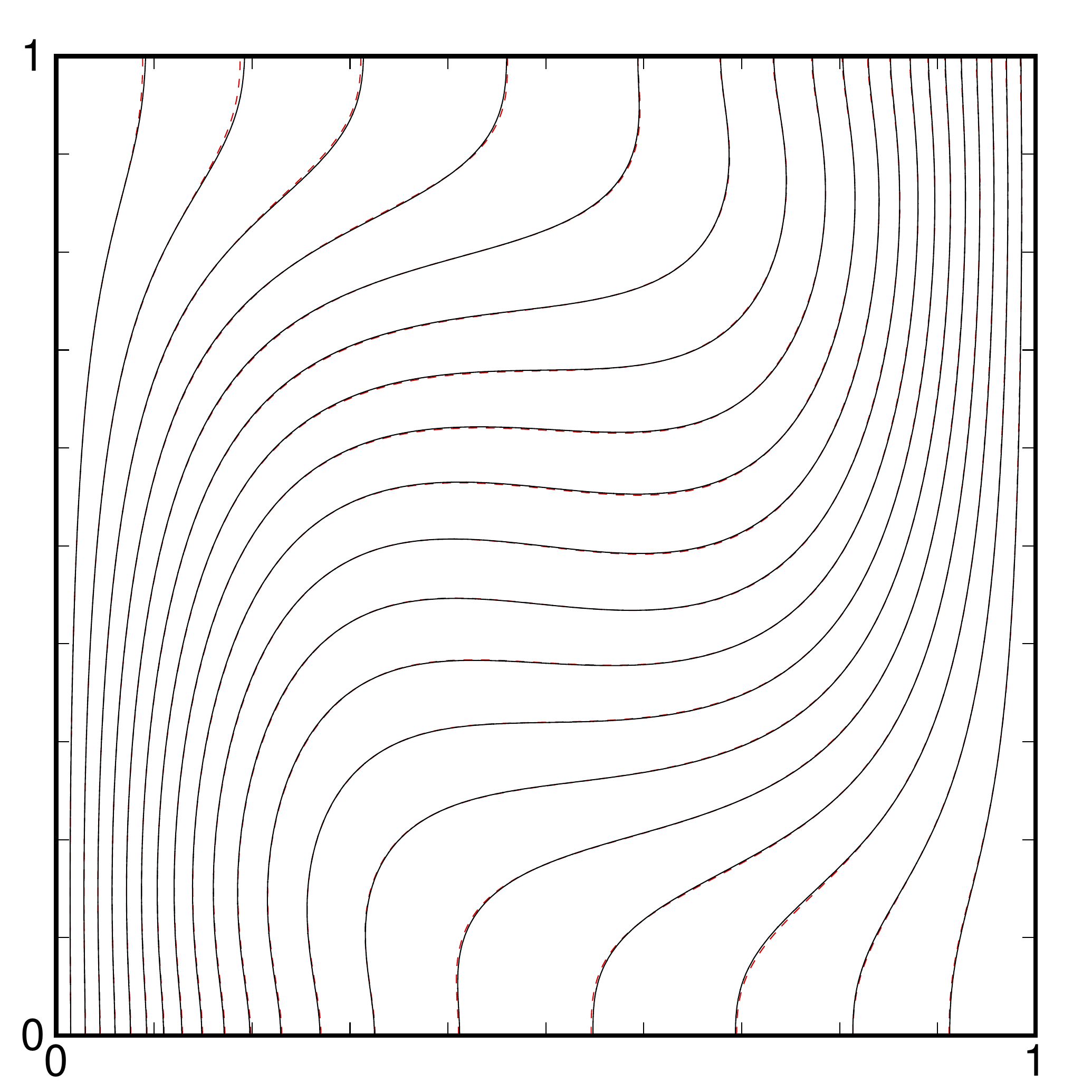}
      \end{subfigure}
        ~
        \begin{subfigure}[b]{0.32\textwidth}
          \includegraphics[width=\textwidth]{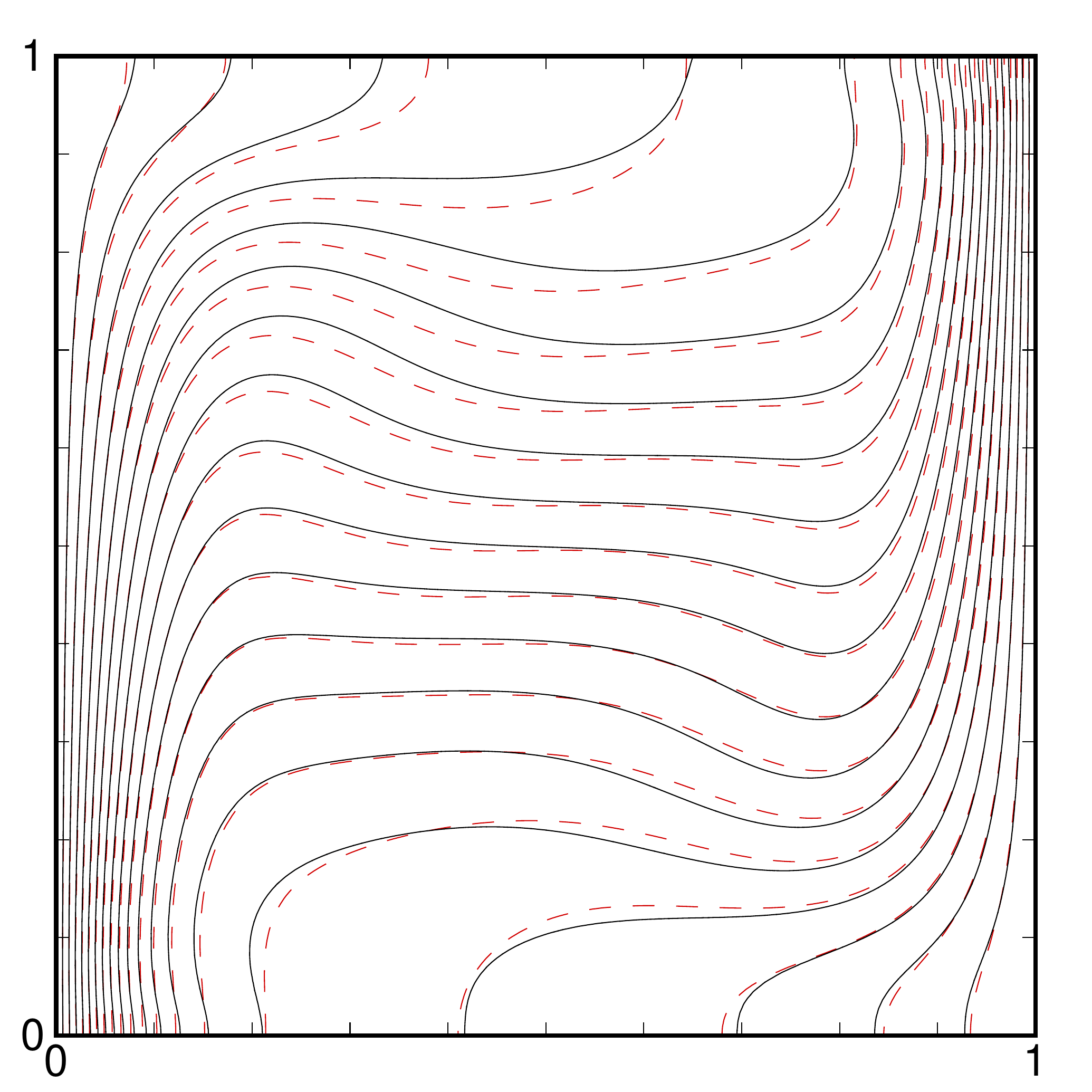}
      \end{subfigure}

	\end{center}
	\caption{Temperature contours for the square cavity test. The high fidelity solution obtained with high fidelity discontinuous Galerkin solver between $1$ and $0$ with the increment of $0.05$ for $Ra=10^3, 10^4,10^5$ from left to right shown with the black contours while the red dashed contours are the solution with PINNs.}
	\label{fig.cavityContour}
\end{figure*}

\begin{figure*}[hbt!]
	\begin{center}
      \begin{subfigure}[b]{0.3\textwidth}
			\includegraphics[width=\textwidth]{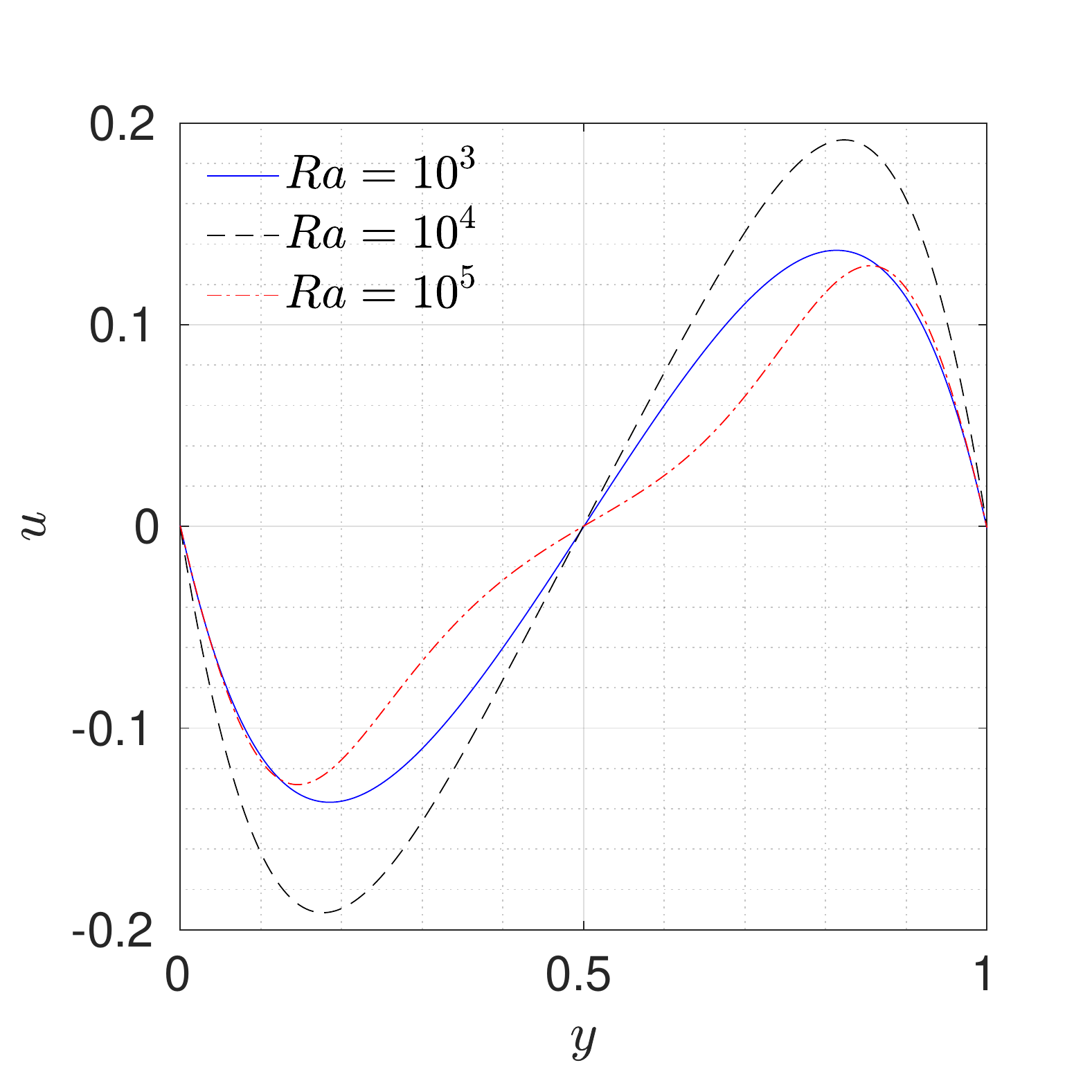} 
		\end{subfigure}
        ~
        \begin{subfigure}[b]{0.3\textwidth}
          \includegraphics[width=\textwidth]{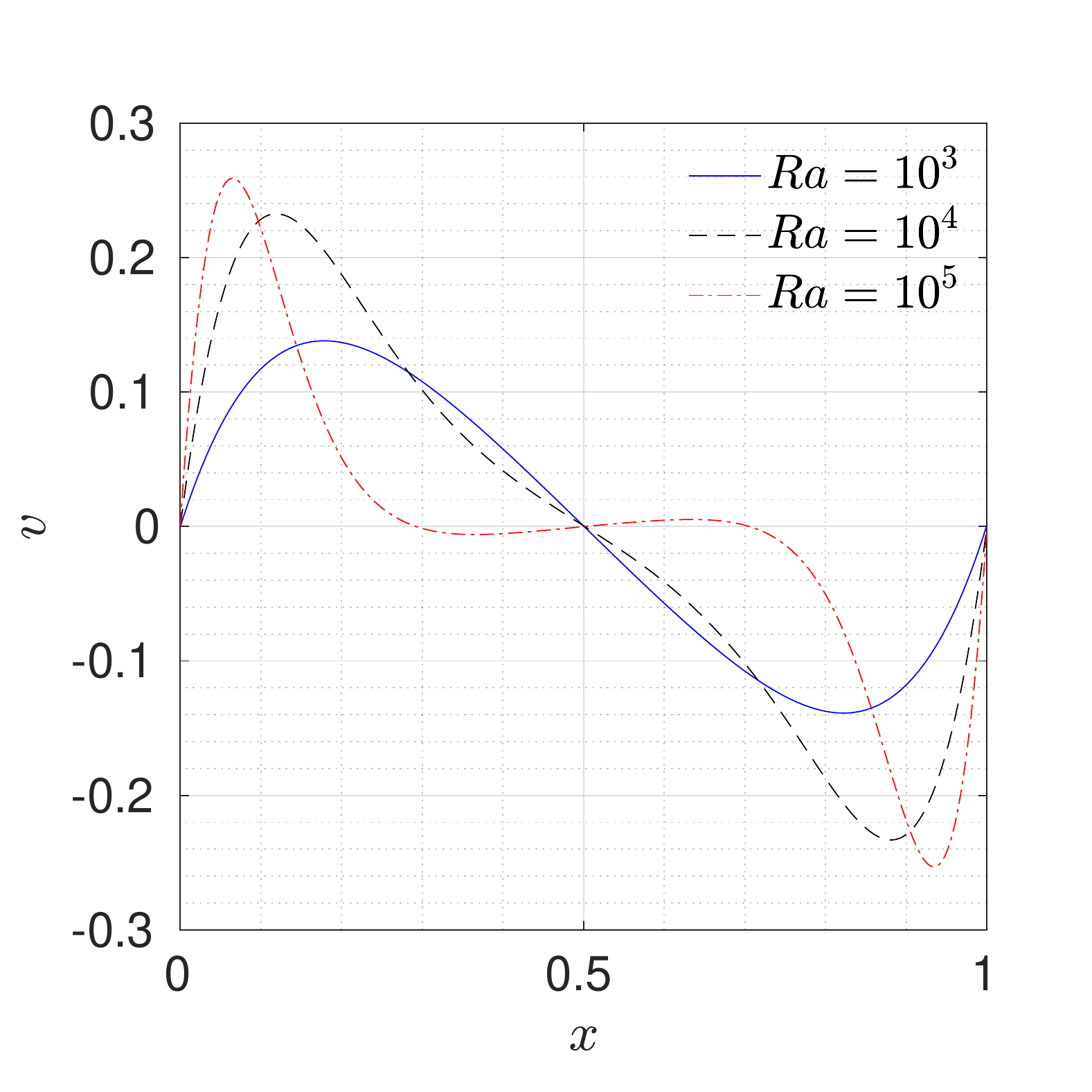}
      \end{subfigure}
        ~
        \begin{subfigure}[b]{0.3\textwidth}
          \includegraphics[width=\textwidth]{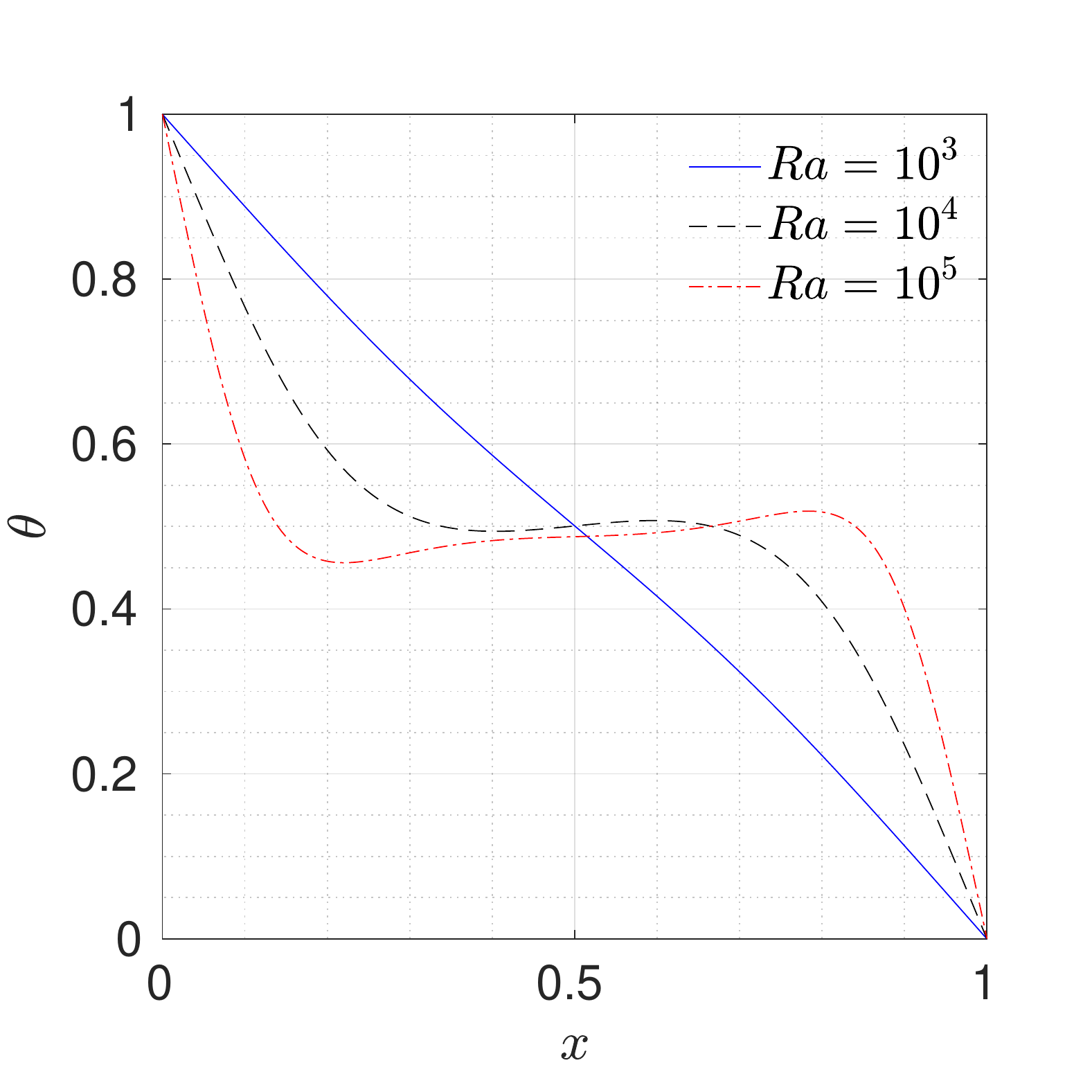}
      \end{subfigure}
      ~
        \begin{subfigure}[b]{0.3\textwidth}
          \includegraphics[width=\textwidth]{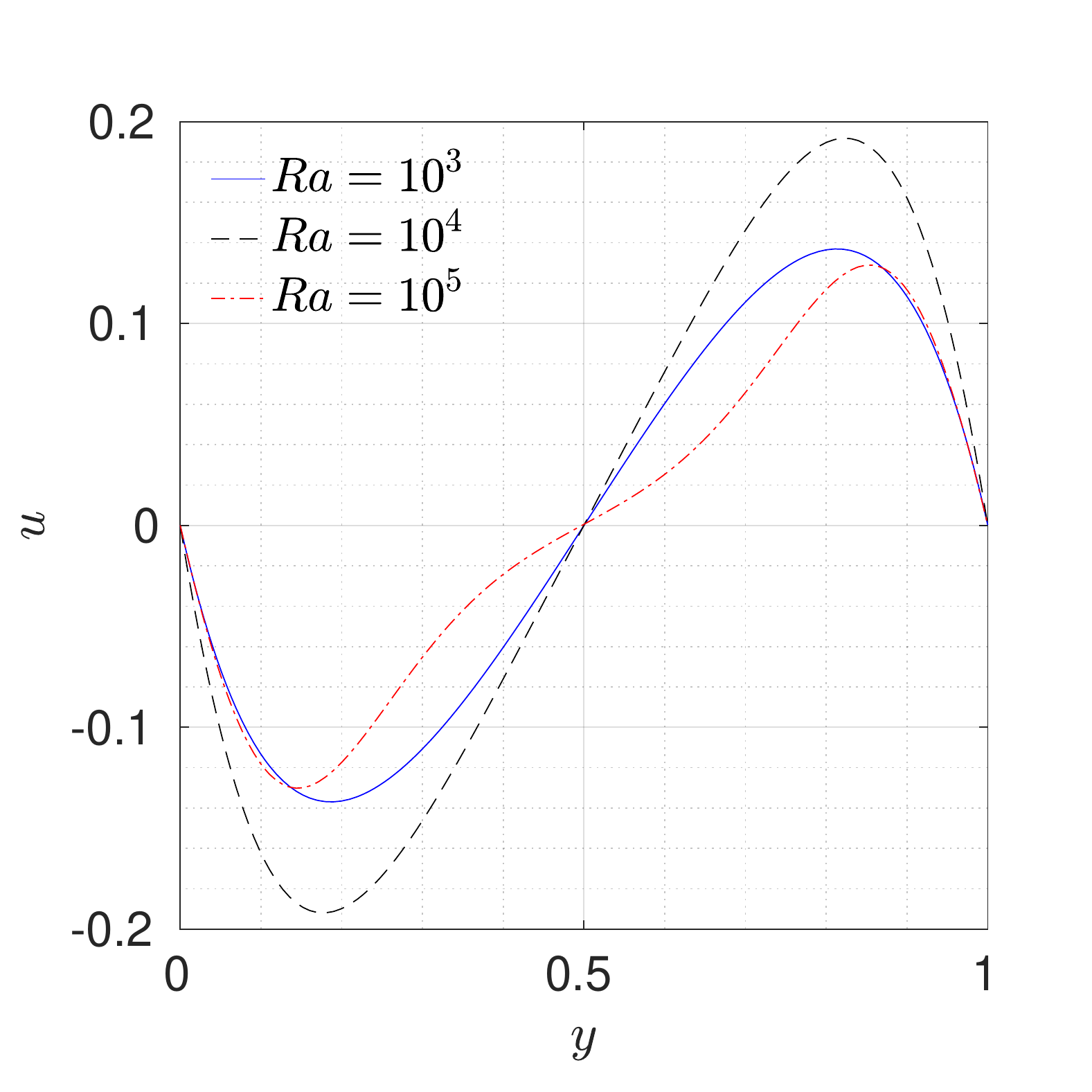}
      \end{subfigure}
      ~
        \begin{subfigure}[b]{0.3\textwidth}
          \includegraphics[width=\textwidth]{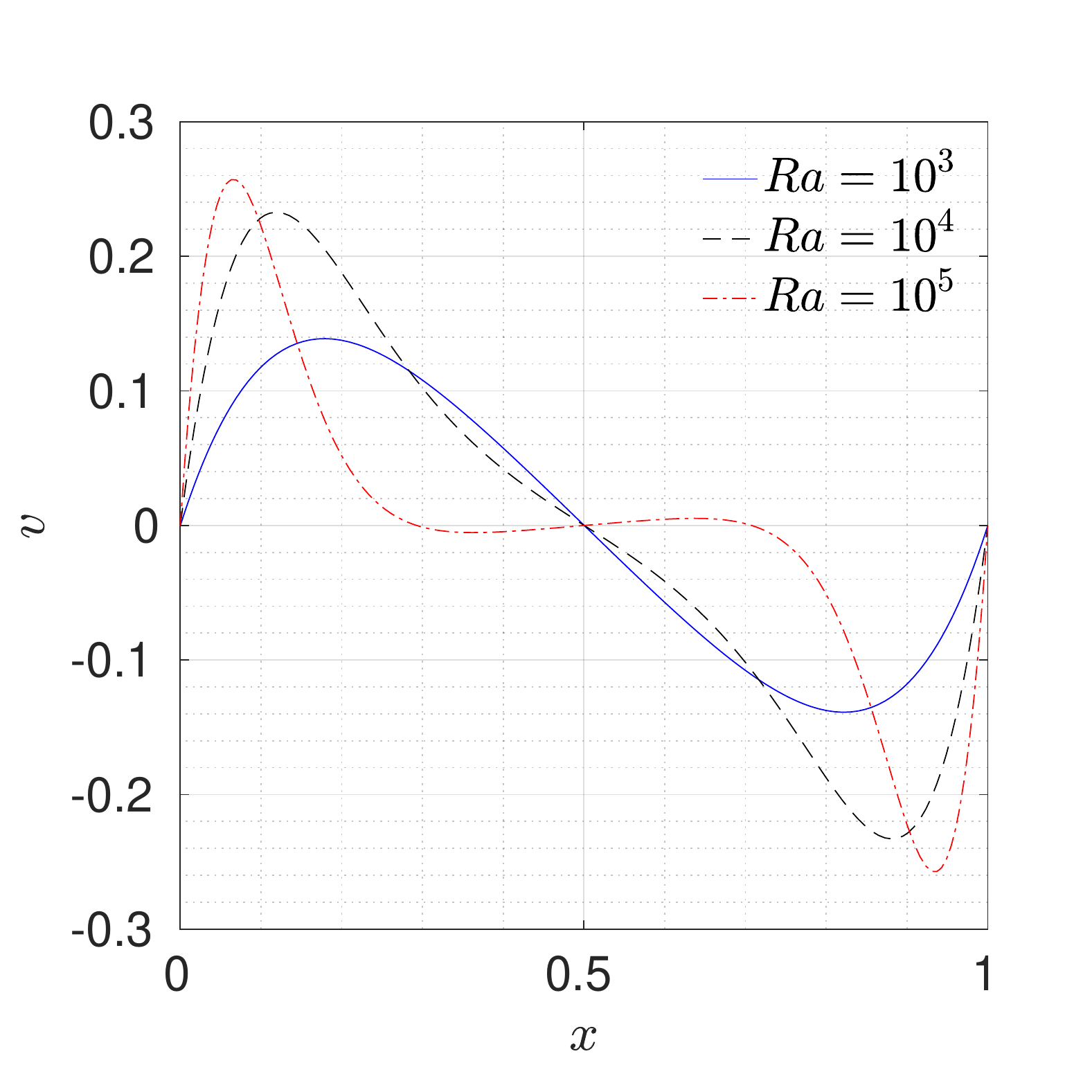}
      \end{subfigure}
      ~
        \begin{subfigure}[b]{0.3\textwidth}
          \includegraphics[width=\textwidth]{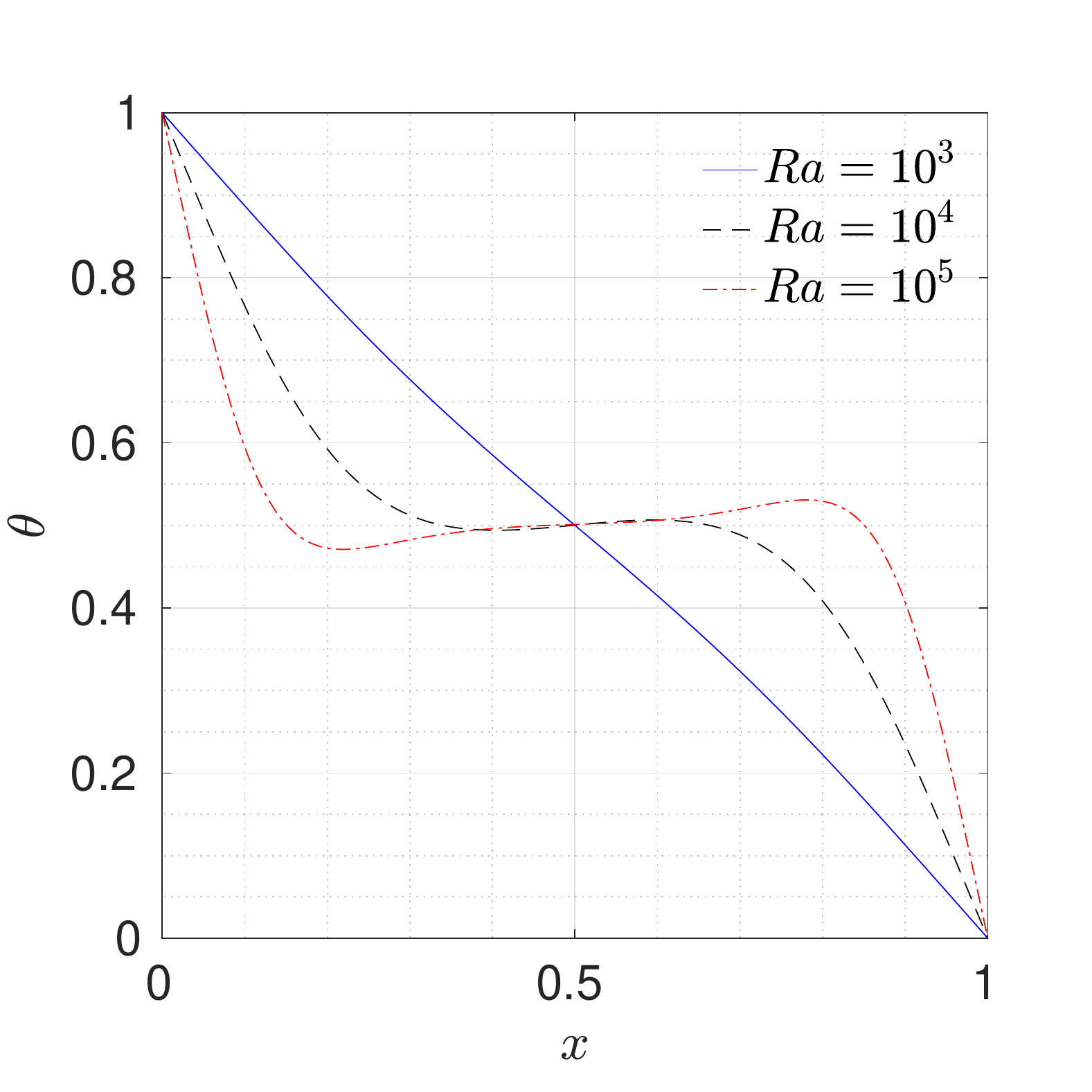}
      \end{subfigure}
	\end{center}
	\caption{Velocity and temperature profiles along the $y=0.5$ and $x=0.5$ lines for different $Ra$ numbers. The first row shows the values obtained with the PINN, while the second row shows the values of high order discontinuous Galerkin solver.}
	\label{fig.cavityProfile}
\end{figure*}

 After the training process, prediction is performed on a $(251\times251)$ grid. In Table \ref{table:square_cavity_1}, we presented the maximum and minimum velocities on the horizontal and vertical centerlines after the prediction with PINN. For cases with different $Ra$ numbers, our framework have values that are comparable with the ones in the literature. In Figure \ref{fig.cavityContour}, the solution of PINN and its comparison with a high fidelity solver through the temperature contours can be seen. Also in Figure \ref{fig.cavityProfile}, the center line profiles of velocity and temperature for different $Ra$ numbers are shown. These contours and profiles qualitatively match with the high order solutions \citep{karakus_accelerated_2022}. To increase the accuracy, we changed the weights of different loss terms as $Ra$ changes. In Table \ref{table:cavity_loss_ratio}, we presented the center line velocities for different $Ra$ numbers and different weight ratio of the residual loss over the boundary loss where $\omega_R$ represents the weight of the residual loss, and $\omega_{BC}$ represents the weight of loss on the boundary conditions. We stopped changing weights when we match the center line velocities with the reference solutions. As the $Ra$ increases, the convective effects inside the domain becomes more dominant. Hence we need to decrease the weight of the boundary losses, and focus more on the residual inside the domain. We select the loss ratio according to Table \ref{table:cavity_loss_ratio} that minimizes the error both inside the domain and on the boundaries.
\begin{table*}[htb!]
    \caption{Maximum and minimum velocities along the center lines with different weight ratio of residual loss and the boundary loss.}
    \centering
    \label{table:cavity_loss_ratio}
    \begin{tabular}{ c c c c c c c c}
         \hline \hline
	\multirow{2}{*}{$\omega_R / \omega_{BC}$}	&       &  \multicolumn{2}{ c }{$ Ra=10^3$} & \multicolumn{2}{c}{$Ra=10^4$} &  \multicolumn{2}{c}{$Ra=10^5$} \\ \cline{3-4} \cline{5-6}  \cline{6-8} 
        &  &  $u_{max}$  &  $v_{max}$       &   $u_{max}$  &  $v_{max}$  &   $u_{max}$  &  $v_{max}$        \\  \hline\hline
		0.5 &     & $0.137 $ & $0.138$ &$0.190$ & $0.231$    & $ 0.137$ &  $0.273$ \\
	 1 &    &  &   &$0.192$ & $0.233$    & $ 0.128$ &  $0.258$     \\
     2&     &  &   & &     & $0.132$ &  $0.261$     \\
	 4&     &  &   & &     & $ 0.130$ &  $0.261$     \\
     \hline \hline
    \end{tabular}
\end{table*}

\begin{figure*}[hbt!]
	\begin{center}
      \begin{subfigure}[b]{0.8\textwidth}
			\includegraphics[width=\textwidth]{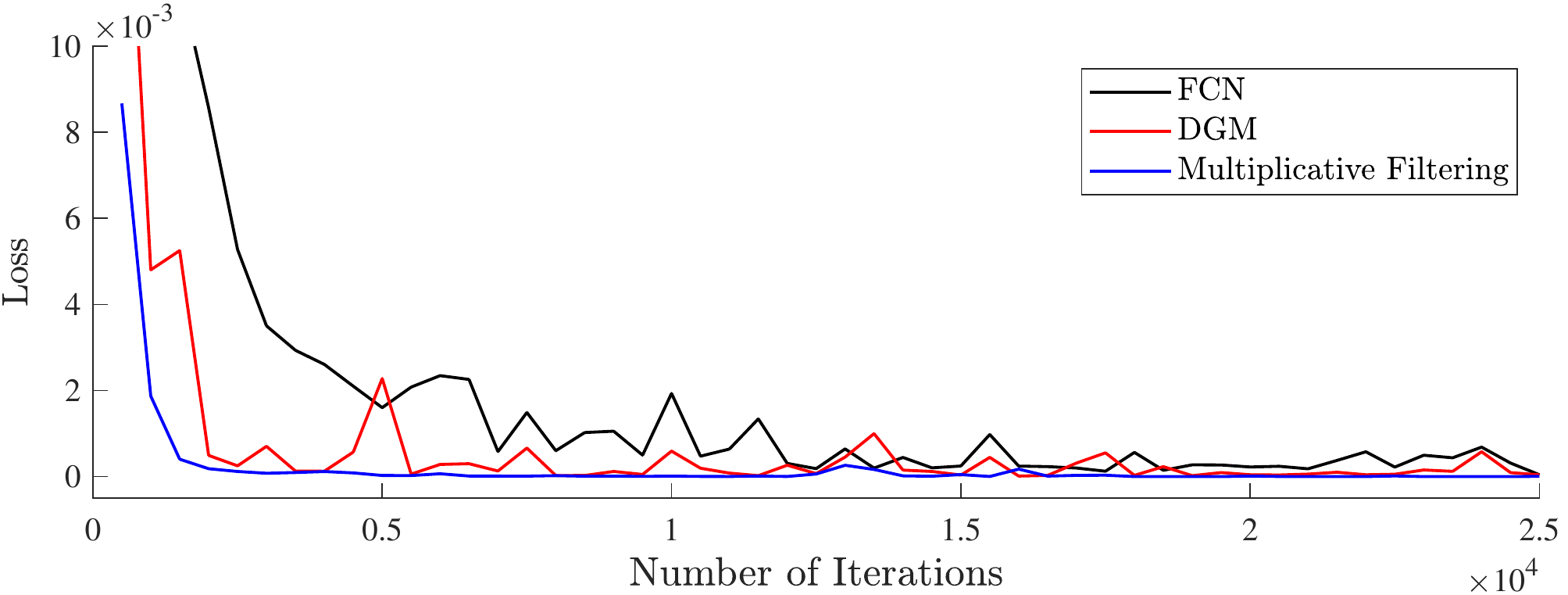} 
		\end{subfigure}
        ~
        \begin{subfigure}[b]{0.8\textwidth}
          \includegraphics[width=\textwidth]{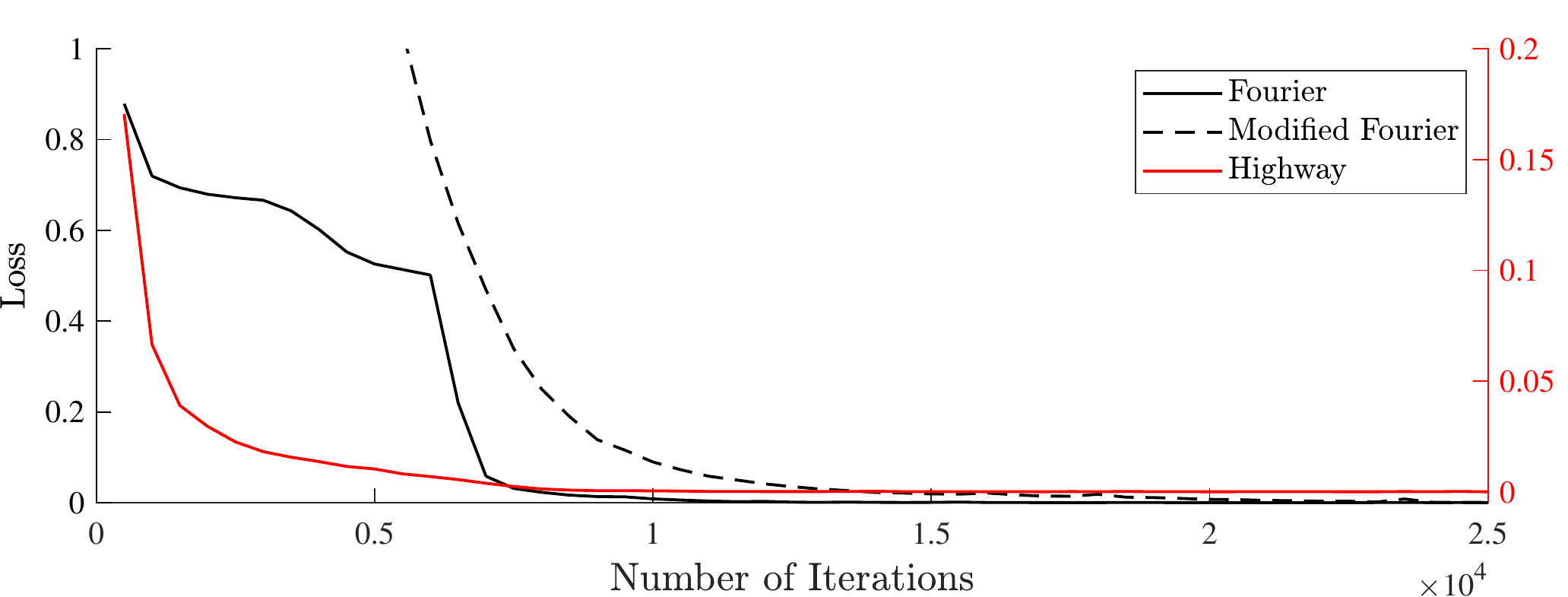}
      \end{subfigure}
	\end{center}
	\caption{Behavior of the total loss in the square cavity problem with $Ra=10^3$ with different types of neural network architectures.}
	\label{fig.numericalTest}
\end{figure*}

We test different types of neural network architectures and monitored the behavior of the total loss of the Adam optimizer for the cavity problem with $Ra=10^3$ and presented in Figure \ref{fig.numericalTest}. The plain fully connected network (FCN), a variation of the fully connected network named as Deep Galerkin Method (DGM) \citep{Sirignano2018DGM}, a Fourier network and a modified Fourier network \citep{wang2021GradientFlow}, a modified highway network using Fourier features \citep{srivastava_training_2015}, and a multiplicative filter network \citep{fathony2021multiplicative} are used. All of these architectures are readily available in NVIDIA Modulus framework. In all the tests 8 layer networks are constructed with 40 units. Hyperbolic tangent is set as the activation function and the learning rate is $10^{-3}$. The architectures that use Fourier mapping converges later than the plain fully connected network since the problem does not have multi-scale behavior. For this simple problem we do not need a Fourier mapping, hence networks that are basically build on plain fully connected networks converges in less iterations. 

\subsection*{Heated Block}
\label{sec.HeatedBlock}
\begin{figure}[hbt!]
    \centering
    \includegraphics[width=\linewidth]{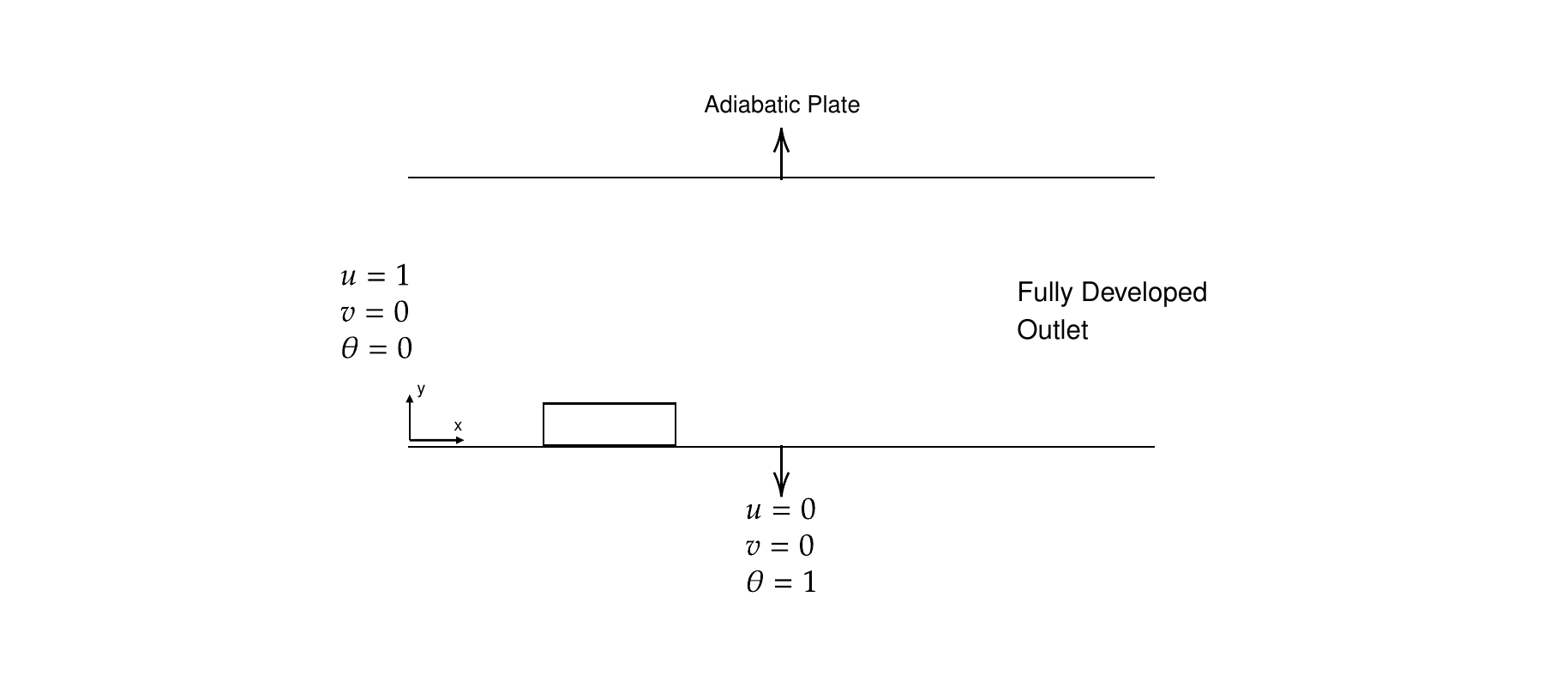}
    \caption{Schematic of the partially blocked channel}
    \label{fig:HeatedBlock}
\end{figure}

In this section we focus on an application of coupled heat transfer with a heat transfer in a partially blocked channel. The domain and the boundary conditions can be seen in Figure \ref{fig:HeatedBlock}. The heated block represents an electronic part on a vertical electronic board \citep{habchi_laminar_1986}. The top wall is adiabatic and the bottom wall is at a prescribed temperature. A low temperature flow comes from the inlet and the outflow is a fully developed outlet meaning the changes in the $x$ direction is zero. The Prandtl number is set to 0.7 for this problem and the Reynolds number is 37.8. The ratio of $Gr/Re^2$ is 1 and the forcing is on the $x$ direction. 

\begin{figure*}[hbt!]
	\begin{center}
        \begin{subfigure}[b]{1.0\textwidth}
          \includegraphics[width=\textwidth]{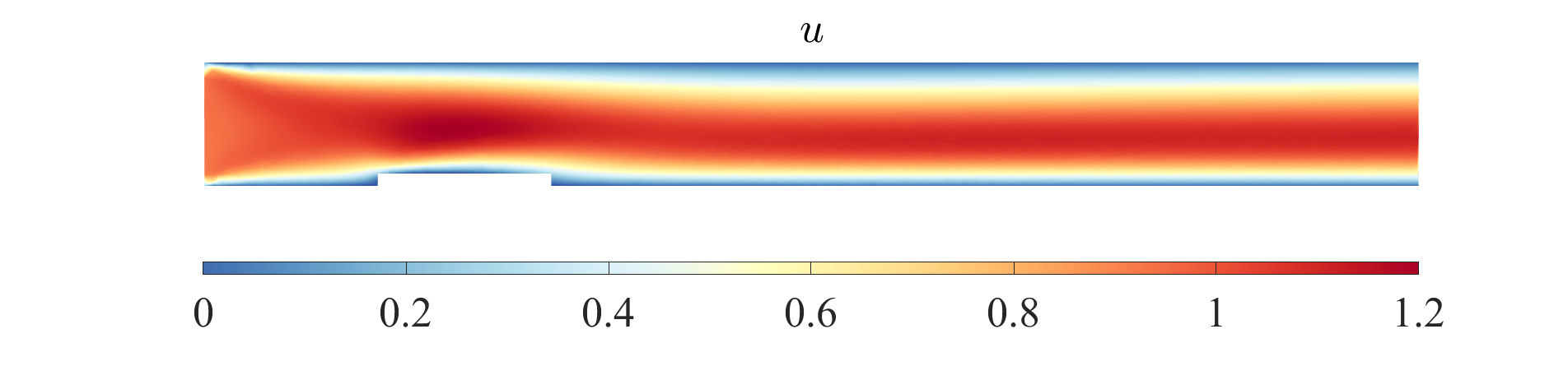}
      \end{subfigure}
      ~
        \begin{subfigure}[b]{1.0\textwidth}
          \includegraphics[width=\textwidth]{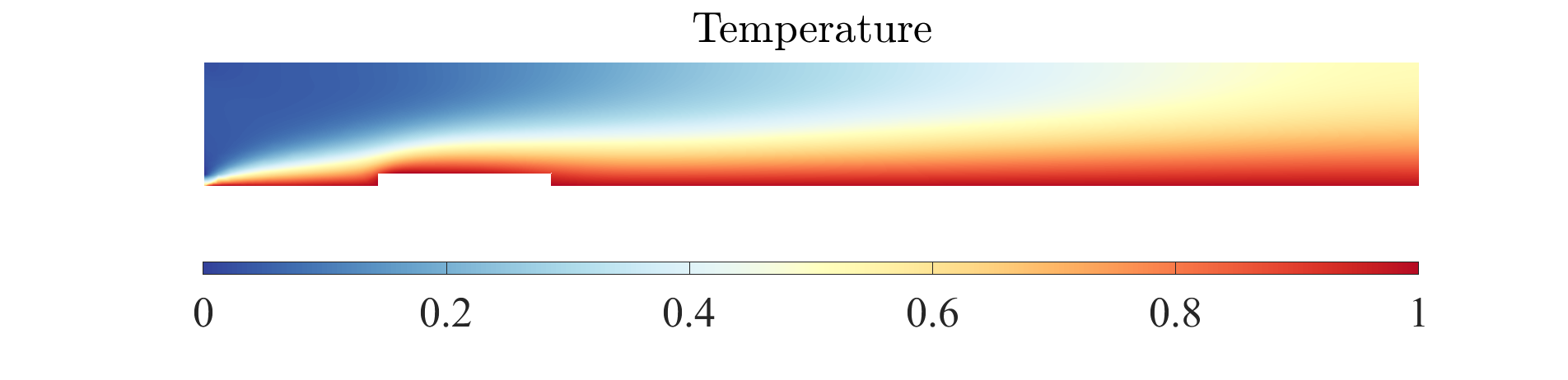}
      \end{subfigure}
	\end{center}
	\caption{Velocity and temperature profiles for the heated block case. The figure on the top shows the velocity profile and the figure on the bottom shows the temperature field predicted by the PINN.}
	\label{fig.heatedBlock}
\end{figure*}

For training the network, we sampled 30 points on the inlet and the outlet, 210 points on the adiabatic wall, total of 40 points on the heated block, total of 180 points on the bottom wall and 1400 points inside the domain. We used 25000 iterations for the Adam optimizer with the learning rate of $5\times10^{-4}$ and obtained the solution presented in Figure \ref{fig.heatedBlock}. PINN solution well predicts the Neumann boundary conditions on the top wall and the fully developed outlet, and the no-slip Dirichlet velocity conditions. 

\subsection*{Multiple Heated Blocks}
In this section we focus on a similar problem in which multiple blocks are present. The problem is time dependent and it is solved for a final time of 8 seconds. The geometric representation of the case is presented in Figure \ref{fig.multiHeatedBlock}. The geometric parameters are given such that \(H=1,\: H/w=2.5,\:L/w=25,\:h/w=0.5\) as presented in \cite{Wu1999_obliquePlate}. At time $t=0$, the initial condition is $u=v=\theta=0$ in the domain. There is a uniform inflow with $u=1$, $v=0$ with the temperature of $\theta=0$. The upper and bottom walls have no-slip conditions as the velocity boundary condition, and Neumann temperature boundary condition of $\partial\theta/\partial n=0$. The blocks also have no-slip conditions and temperature boundary condition of $\partial\theta/\partial n=-1$. Gravity is on $y$ direction, and flow parameters are given as $Re=400$, $Pr=0.7$, and $Gr/Re^2=0.5$.

\begin{figure}[hbt!]
    \centering
    \includegraphics[width=\textwidth]{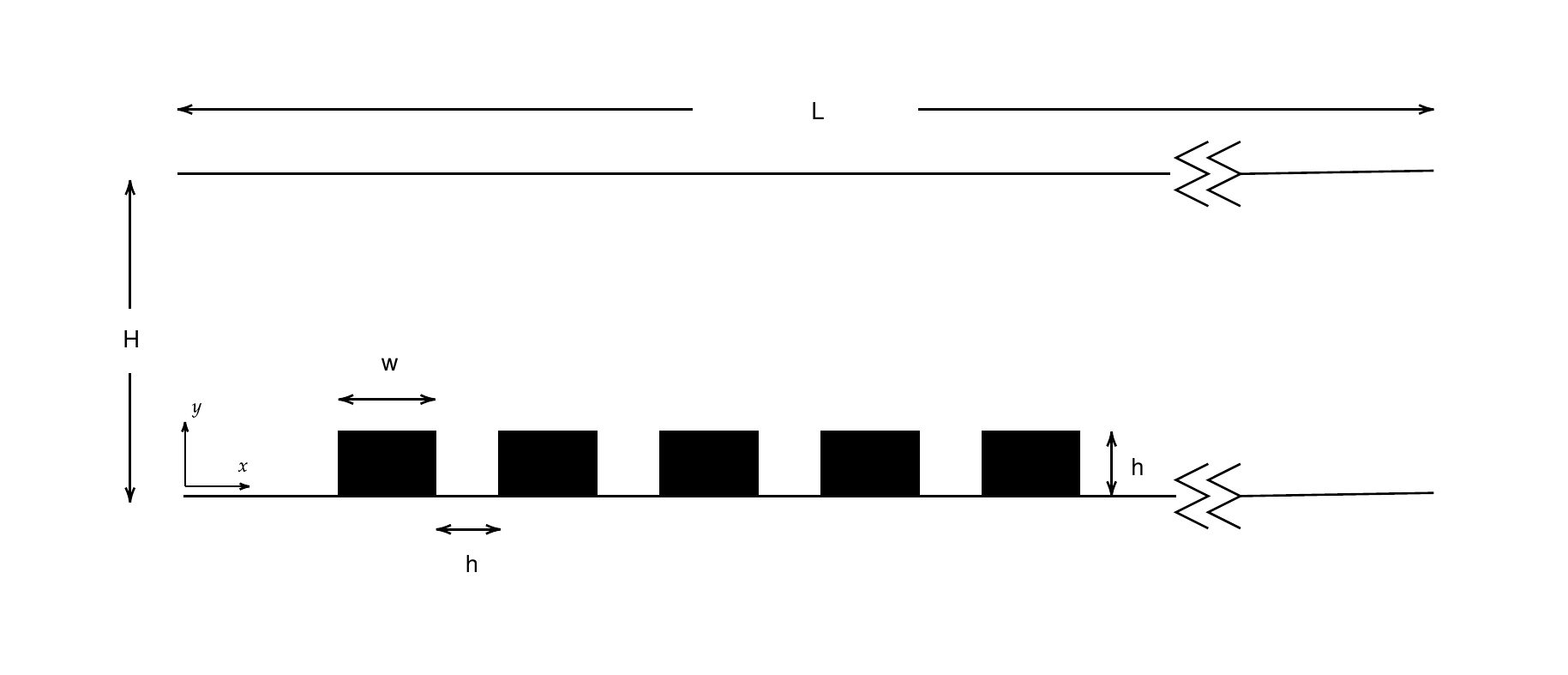}
    \caption{Geometric representation of the channel with multiple blocks}
    \label{fig.multiHeatedBlock}
\end{figure}

In the training phase, 50 points on the inlet and the outlet is generated uniformly. 40 points on each block, 500 points on top wall and 450 points on the bottom wall is sampled. The problem is solved with continuous time approach such that we treat time $t$ as another variable as the spatial coordinates instead of approaching the time sequentially. Adam optimizer is used again to find the optimized hyperparameters with a learning rate of $5\times10^{-4}$. The solution at the final time $t=8s$ is shown in Figure \ref{fig.multiHeatedBlock}. The horizontal velocity and the temperature fields are plotted and it can be seen that PINN solution represents the flow field well physically inside the domain and also satisfy the Dirichlet and Neumann boundary conditions.

\begin{figure*}[hbt!]
	\begin{center}
        \begin{subfigure}[b]{1.0\textwidth}
          \includegraphics[width=\textwidth]{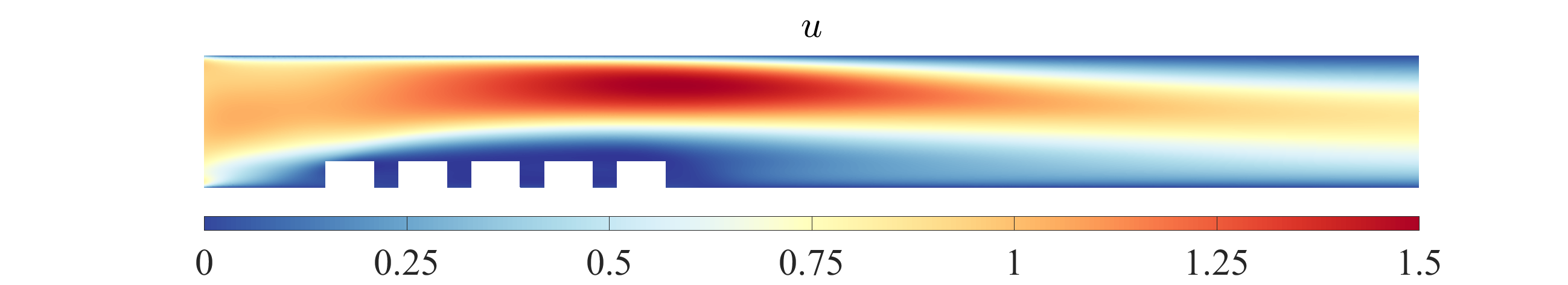}
      \end{subfigure}
      ~
        \begin{subfigure}[b]{1.0\textwidth}
          \includegraphics[width=\textwidth]{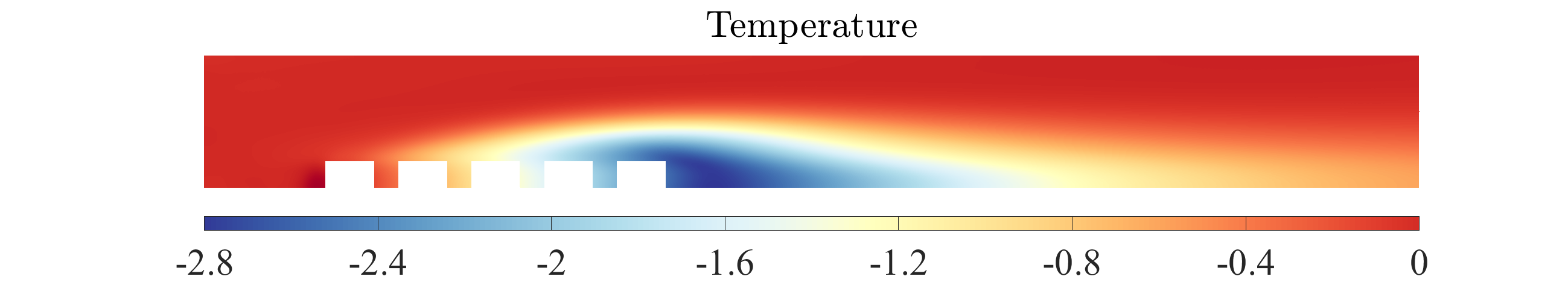}
      \end{subfigure}
	\end{center}
	\caption{Velocity and temperature profiles for the multiple heated blocks case at time $t=8s$. The figure on the top shows the velocity profile and the figure on the bottom shows the temperature field predicted by the PINN.}
	\label{fig.multiHeatedBlock}
\end{figure*}

\section*{CONCLUSION}

In this work, we solve two dimensional incompressible thermal convection problems with physics informed neural networks. Using Boussinesq approximation, the incompressible Navier-Stokes equation is coupled with the energy equation written in terms of temperature. The loss function is constructed using these PDEs and necessary boundary/initial conditions. The PINN solution predicts the flow and temperature fields well compared to the analytical solutions and high order solvers in various problems. We show that adding observations into the flow field increases the accuracy of the prediction and the framework is very sensitive to the weights of the individual loss terms. Furthermore, we consider two different channel problems with partial blockage that resemble power electronics applications. The first one is a simple steady problem and the other is an unsteady problem in which we tackle it with a continuous approach such that the time is treated as another variable and we obtained results that resembles the flow physics well. For a future work, the time dependent problems can be implemented with time marching approaches such as Recurrent Neural Networks or Gated Response Units.


\bibliographystyle{apalike}
\bibliography{main}

\end{document}